\documentclass[journal,12pt,draftclsnofoot,onecolumn]{IEEEtran}

\usepackage[cmex10]{amsmath}
\usepackage{amssymb}

\usepackage{cases}
\usepackage{multirow}
\usepackage{empheq}

\usepackage{cite}
\usepackage{verbatim}

\usepackage{algorithm}
\usepackage{algpseudocode}

\ifCLASSINFOpdf
  \usepackage[pdftex]{graphicx}
\else
  \usepackage[dvips]{graphicx}
\fi
\usepackage[caption=false,font=footnotesize]{subfig}

\usepackage{color}
\usepackage{multirow}

\newtheorem{theorem}{Theorem}
\newtheorem{lemma}[theorem]{Lemma}

\newtheorem{corollary}[theorem]{Corollary}
\newtheorem{definition}[theorem]{Definition}

\newtheorem{proposition}[theorem]{Proposition}
\newtheorem{remark}[theorem]{Remark}

\newtheorem{claim}[theorem]{Claim}

\newcommand{\vect}[1]{\mathbf{#1}}

\begin{document}

\sloppy

\title{Generalized Water-filling for Source-aware Energy-efficient SRAMs}

\author{
\IEEEauthorblockN{Yongjune Kim, Mingu Kang, Lav R. Varshney, and Naresh R. Shanbhag} \\
\IEEEauthorblockA{Coordinated Science Laboratory, 
	University of Illinois at Urbana--Champaign\\
    Urbana, IL, USA\\
    Email: \{yongjune, mkang17, varshney, shanbhag\}@illinois.edu}

\thanks{This work was supported in part by Systems on Nanoscale Information fabriCs (SONIC), one of the six SRC STARnet Centers, sponsored by MARCO and DARPA.}
}

%\author{\IEEEauthorblockN{Yongjune Kim\IEEEauthorrefmark{1},
%	Abhishek A. Sharma\IEEEauthorrefmark{1},		
%	Robert Mateescu\IEEEauthorrefmark{2},
%	Seung-Hwan Song\IEEEauthorrefmark{2},
%	Zvonimir Z. Bandic\IEEEauthorrefmark{2},\\
%	James A. Bain\IEEEauthorrefmark{1},
%	and
%	B. V. K. Vijaya Kumar\IEEEauthorrefmark{1}}
%\IEEEauthorblockA{\IEEEauthorrefmark{1}Data Storage Systems Center (DSSC), Carnegie Mellon University, Pittsburgh, PA, USA\\ Email: \{yongjunekim, abhisheksharma\}@cmu.edu, \{jbain, kumar\}@ece.cmu.edu}
%\IEEEauthorblockA{\IEEEauthorrefmark{2}HGST Research, San Jose, CA, USA\\
%Email: \{robert.mateescu, seung-hwan.song, zvonimir.bandic\}@hgst.com}
%}

%% To balance the two columns, you should reduce the text-height of
%% the last page using the following command:
%%%%%%%%%%%%%%%%%%%%%%%%%%%%%%%%%%%%%%%%%%%%%%%%%%%%%%%%%%%%%%%%%%%%%
%\addtolength{\textheight}{-9.35cm}
%%%%%%%%%%%%%%%%%%%%%%%%%%%%%%%%%%%%%%%%%%%%%%%%%%%%%%%%%%%%%%%%%%%%%
%% with an appropriate value. This command must be place on the second
%% last page, i.e., for a one-page abstract here, for a two-page
%% abstract right after the \maketitle command.

%% Create the title:
\maketitle

\begin{abstract}
Conventional low-power static random access memories (SRAMs) reduce read energy by decreasing the bit-line voltage swings uniformly across the bit-line columns. This is because the read energy is proportional to the bit-line swings. On the other hand, bit-line swings are limited by the need to avoid decision errors especially in the most significant bits. We propose a principled approach to determine optimal non-uniform bit-line swings by formulating convex optimization problems. For a given constraint on mean squared error of retrieved words, we consider criteria to minimize energy (for low-power SRAMs), maximize speed (for high-speed SRAMs), and minimize energy-delay product. These optimization problems can be interpreted as classical water-filling, ground-flattening and water-filling, and sand-pouring and water-filling, respectively. By leveraging these interpretations, we also propose greedy algorithms to obtain optimized discrete swings. Numerical results show that energy-optimal swing assignment reduces energy consumption by half at a peak signal-to-noise ratio of 30dB for an 8-bit accessed word. The energy savings increase to four times for a 16-bit accessed word. 
\end{abstract}

\section{Introduction}

Von Neumann computing architectures separate memory units from computing units so there is frequent data access that consumes enormous energy. Since static random access memories (SRAMs) access requires more energy than arithmetic operations~\cite{Horowitz2014computing}, SRAM access energy accounts for the significant part of the total energy consumption in many information processing circuits~\cite{Lin2006mpeg4,Sinangil2014application,Han2016eie,Chen2017eyeriss,Sze2017efficient}. Thus, it is important to reduce the energy consumption of SRAM access. The basic way to reduce the access energy is to decrease either supply voltages or bit-line (BL) swings, which increases vulnerability to variations and noise. If we reduce supply voltages or BL swings across all BL columns~\cite{Zhai2007sub,Abu-Rahma2010reducing}, then bit error rates (BERs) of all bit positions increase equally.

In many applications including signal processing and machine learning (ML) tasks, however, the impact of bit errors depends on bit position. For example, errors in the most significant bits (MSBs) of image pixels degrade overall image quality much more than errors in the least significant bits (LSBs). Likewise, an MSB error can cause a catastrophic loss in the inference accuracy of ML applications. 

Until now, the following techniques have been proposed to address the different impacts of each bit position for energy efficiency: 
\begin{enumerate}
	\item Storing the MSBs in more robust bit cells and the LSBs in less robust cells~\cite{Chang2011priority,Kwon2012heterogeneous}, 
	\item Applying higher supply voltage for the MSBs and lower supply voltage for the LSBs~\cite{George2006probabilistc,Yi2008partial,Cho2011reconfigurable},
	\item Unequal error protection (UEP) by error control codes (ECCs)~\cite{Yang2011unequal,Tang2016unequal},
	\item LSB dropping (dropping the LSBs at the cost of reduced arithmetic precision)~\cite{Kaul2012variable,Frustaci2015sram,Frustaci2016approximate}.
\end{enumerate}

The first approach requires costly bit cells redesign and manual array reorganization. Also, the bit cells are fixed at design time, so it is unable to dynamically track the time-varying fidelity requirement~\cite{Frustaci2016approximate}. The second approach employs different supply voltages for each bit position, which significantly complicates the power routing network. Practical implementations only allow a few supply voltage levels~\cite{Yi2008partial,Cho2011reconfigurable}. Fine-grained UEP~\cite{Yang2011unequal,Tang2016unequal} requires complicated hardware implementations and dynamic change of protection is limited. LSB dropping~\cite{Kaul2012variable,Frustaci2015sram,Frustaci2016approximate} enables dynamic fidelity control by changing the number of dropped LSBs. Note that UEP and LSB dropping allow two levels of granularity (protected/unprotected or dropped/undropped) for each bit position.  

In~\cite{Frustaci2015sram,Frustaci2016approximate}, selective ECCs were proposed by combining UEP and LSB dropping. Since parity bits are stored in dropped LSB-cells, the encoded data has the same length as the uncoded data. In~\cite{Huang2015acoco}, the authors proposed adaptive coding techniques for different computations on the data read from faulty memories.

This paper presents an information-theoretic approach to determine the optimal BL swing assignments. For a given constraint on mean squared error (MSE) of retrieved words, we formulate convex optimization problems whose objectives are as follows:
\begin{itemize}
	\item[C1.]{Minimize energy (low-power SRAMs),}
	\item[C2.]{Maximize speed (high-speed SRAMs),}
	\item[C3.]{Minimize energy-delay product (EDP).}
\end{itemize}
Solutions to these convex problems yield optimal performance that is theoretically attainable. By casting read access for SRAMs as communication over parallel channels, we investigate the fundamental trade-offs between physical resources (energy, delay, and EDP) and a fidelity (MSE) constraint.

%The optimal swings for low-power SRAMs reduce the energy consumption by half at a peak signal-to-noise ratio (PSNR) of 30dB for an 8-bit accessed word. The energy savings increase to four times for a 16-bit accessed word.  

In addition, we provide \emph{generalized water-filling} interpretations for our optimal solutions. This follows since accessing a $B$-bit word is equivalent to communicating information through $B$ parallel channels. In classical water-filling, the ground represents the noise levels of parallel channels~\cite{Shannon1949communication,Cover2006}. On the other hand, the importance of each bit position determines the ground level in our optimization problems. Each optimization problem has its own interpretation depending on its objective function: water-filling (C1), ground-flattening and water-filling (C2), and sand-pouring and water-filling (C3), respectively. We also observe interesting connections between our problems and variants on water-filling such as \emph{constant-power water-filling}~\cite{Chow1993bandwidth,Yu2001constant} and \emph{mercury/water-filling}~\cite{Lozano2006optimum}. Also, we show that the proposed optimization techniques can be extended to a wide range of sources and noise models.

Furthermore, we propose an SRAM circuit architecture to assign non-uniform bit-level swings. The proposed architecture separates the data for each bit position in different SRAM subarrays by interleaving. The proposed architecture enables fine-grained and dynamic control of bit-level swings depending on time-varying fidelity requirements with little circuit complexity overhead. Also, we propose greedy algorithms to optimize swing values drawn from a discrete set due to circuit implementation limitations. Generalized water-filling interpretations and Karush-Kuhn-Tucker (KKT) conditions are leveraged to develop these discrete optimization algorithms.   

The rest of this paper is organized as follows. Section~\ref{sec:basics} introduces key metrics of energy, delay, and fidelity. Section~\ref{sec:proposed} formulates the convex optimization problems to determine the optimum bit-level swings and provides generalized water-filling interpretations. Section~\ref{sec:extension} shows that the proposed optimization techniques can be extended to various source and noise models. Section~\ref{sec:discrete} investigates the SRAM architecture and develops greedy algorithms to optimize discrete swings. Section~\ref{sec:numerical} gives numerical results and Section~\ref{sec:conclusion} concludes.

%In many applications including machine learning and media (image, audio, and video) processing, a certain amount of error can be tolerated since those applications involve computation that is statistical in nature. By exploiting this statistical behavior, aggressive energy savings can be achieved without severely compromising the correctness of the overall computation~\cite{Shanbhag2010stochstic,Han2013approximate,George2006probabilistc,Frustaci2016approximate}. 

\section{SRAM Metrics for Resource and Fidelity}\label{sec:basics}

\begin{figure}[!t]
	\centering
	\vspace{-3mm}
	\includegraphics[width=0.40\textwidth]{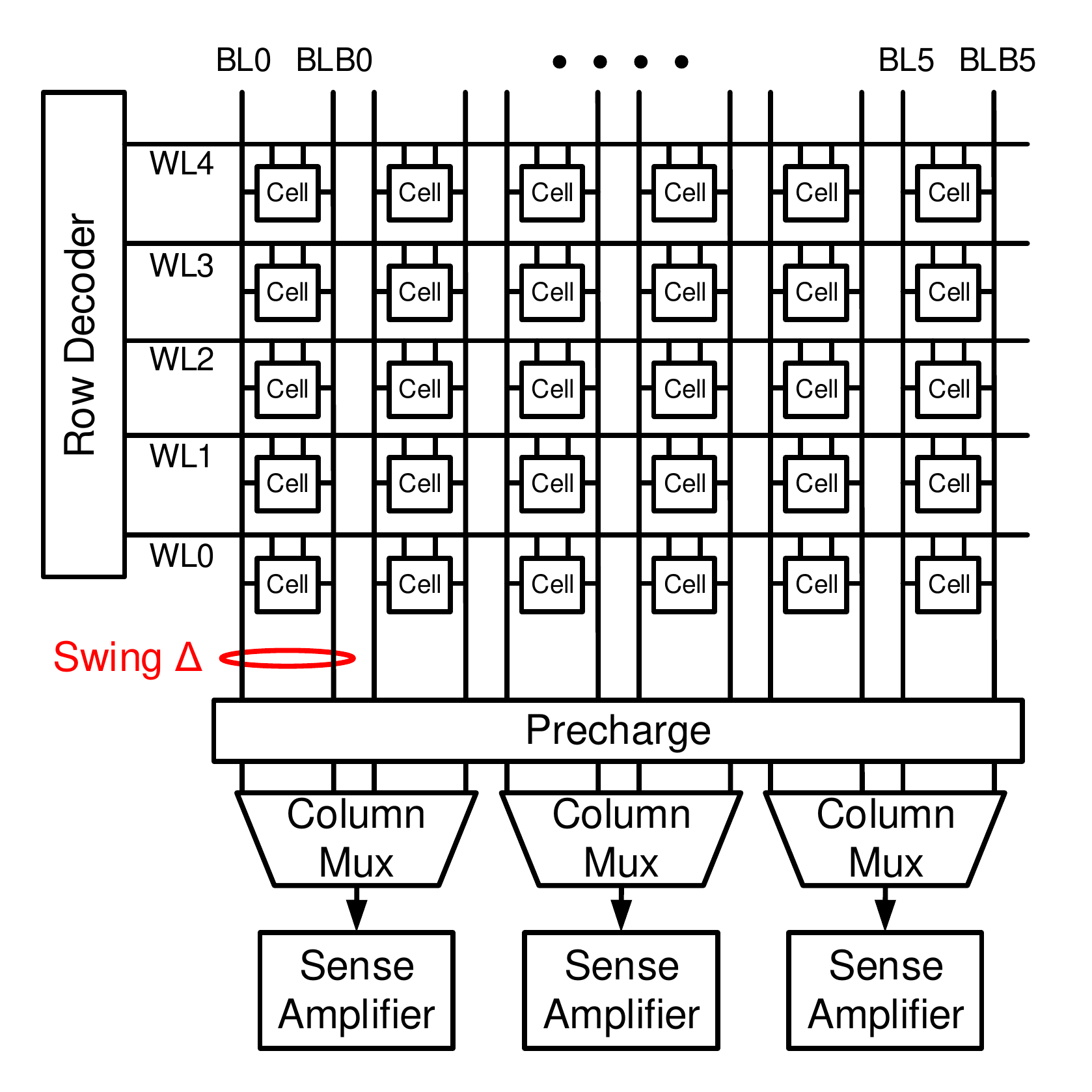}
	\vspace{-9mm}
	\caption{A typical $N_\text{BL} \times N_\text{WL}$ SRAM block ($N_\text{BL}=6$ and $N_\text{WL}=5$).}
	\label{fig:sram}
	\vspace{-8mm}
\end{figure}

The total energy in an SRAM read access is given by
\begin{equation}\label{eq:energy_total}
\mathsf{E}_{\text{total}} = \mathsf{E}_{\text{array}} + \mathsf{E}_{\text{peri}} + \mathsf{E}_{\text{leakage}}  
\end{equation}
where $\mathsf{E}_{\text{array}}$ and $\mathsf{E}_{\text{peri}}$ denote the dynamic energy consumption from the SRAM bit cell array and the peripheral circuitry, respectively, and $\mathsf{E}_{\text{leakage}}$ represents the energy loss due to leakage. $\mathsf{E}_{\text{array}}$ is the dominant component of energy consumption in high-density SRAMs during normal read operations~\cite{Abu-Rahma2010reducing,Macii2002memory,Sinangil2014application}. Hence, we focus on $\mathsf{E}_{\text{array}}$, which is given by
\begin{align}
\mathsf{E}_{\text{array}} &\propto N_{\text{BL}} N_{\text{WL}} C_{\text{bit}}  V_{\text{dd}} \Delta \label{eq:energy_array}
\end{align}
where $N_{\text{BL}}$ and  $N_{\text{WL}}$ are the numbers of bit-lines (BLs) and word-lines (WLs) in a memory bank, respectively. $C_{\text{bit}}$ is the BL capacitance per bit cell and $V_{\text{dd}}$ is the supply voltage. Also, $\Delta$ denotes the BL voltage swing in read access. As shown in Fig.~\ref{fig:sram}, the voltage swing $\Delta$ is the voltage difference between BL and BL-bar (BLB). This voltage difference occurs because either BL or BLB is discharged according to the stored bit. A sense amplifier detects which line (BL or BLB) has the higher voltage and decides whether the corresponding bit cell stores 1 or 0.

The swing $\Delta$ can be controlled by changing the WL pulse-width (i.e., WL activation time) $T_{\text{WL}}$ since 
\begin{equation} \label{eq:delta_pulsewidth}
\Delta = \frac{I_c}{N_{\text{WL}} C_{\text{bit}}} \cdot T_{\text{WL}}
\end{equation}
where $I_c$ is the discharge current of BL corresponding to the accessed bit cell~\cite{Abu-Rahma2010reducing}. From \eqref{eq:energy_array} and \eqref{eq:delta_pulsewidth}, we can observe that $\mathsf{E}_{\text{array}}$ is directly proportional to $T_{\text{WL}}$. Also, $T_{\text{WL}}$ has a direct impact on the read access time~\cite{Abu-Rahma2010reducing,Abu-Rahma2011sa}.

Since larger voltage swing $\Delta$ improves noise margin, there are trade-off relations between reliability, energy, and delay. These relations will be explained in the following subsections. 

\subsection{Resource Metrics for Accessing $B$-bit Word: Energy, Delay, and EDP}

We define resource metrics for energy, delay, and EDP for accessing a $B$-bit word. First, read access energy can be defined as follows. 
\begin{definition}\label{def:energy}The read energy to access a $B$-bit word is
	\begin{equation}\label{eq:energy}
	\mathsf{E}(\vect{\Delta}) = \sum_{b=0}^{B-1}{\Delta_b} = \vect{1}^{\mathsf{T}} \vect{\Delta}
	\end{equation}
	where $\vect{1}$ denotes the all-one vector and the superscript $\mathsf{T}$ denotes transpose. Note that $\vect{\Delta} = \left(\Delta_0, \ldots, \Delta_{B-1}\right)$ where $\Delta_b$ denotes the swing for the $b$th bit position in a $B$-bit word. Note that $\mathsf{E}(\vect{\Delta})$ represents $\mathsf{E}_{\text{array}}$ in \eqref{eq:energy_total}. 
\end{definition}

\begin{definition}\label{def:max_swing} The maximum swing corresponding to a $B$-bit word is 
	\begin{equation} \label{eq:max_swing}
	\rho = \max(\vect{\Delta}) = \max\left\{\Delta_0, \ldots, \Delta_{B-1}\right\}. 
	\end{equation}
\end{definition}
If we allot non-uniform swings for each bit position, the access time for a $B$-bit word depends on $T_{\text{max}} = \max\{T_{\text{WL},0},\ldots,T_{\text{WL},B-1}\}$ where $T_{\text{WL},b}$ denotes the WL pulse-width for the $b$th bit position. Note that $T_{\text{max}}$ is the pulse-width corresponding to the maximum swing $\rho$ because of \eqref{eq:delta_pulsewidth}. Hence, the maximum swing $\rho$ is a proper metric to be minimized to maximize read speed.

The EDP is considered to be a fundamental metric as it captures the trade-off between energy and delay~\cite{Horowitz1994low,Gonzalez1996energy}. We define the EDP for accessing a $B$-bit word based on Definitions~\ref{def:energy} and \ref{def:max_swing}. 

\begin{definition}\label{def:edp}The EDP to access a $B$-bit word is
	\begin{equation}\label{eq:edp}
	\mathsf{EDP}(\vect{\Delta}) = \mathsf{E}(\vect{\Delta}) \cdot \rho = \vect{1}^{\mathsf{T}} \vect{\Delta} \cdot \rho. 
	\end{equation}	
\end{definition}

\subsection{Fidelity Metric for Accessing $B$-bit Word: MSE}

We will define a fidelity metric for accessing a $B$-bit word. Suppose that a $B$-bit word $x = (x_0, \ldots, x_{B-1})$ is stored in SRAM cells, where $x_0$ and $x_{B-1}$ are the LSB and MSB, respectively. Note that $x$ can be represented by
\begin{equation}\label{eq:representation}
x = \sum_{b=0}^{B-1}{2^b x_b}
\end{equation}
where $x_b \in \left\{0, 1\right\}$ and $x \in\left[0, 2^B -1\right]$ (for integers $i$ and $j$ such that $i<j$, $\left[i,j\right] = \left\{i, \ldots, j\right\}$). Also, $\widehat{x} = (\widehat{x}_0, \ldots, \widehat{x}_{B-1})$ denotes the retrieved $B$-bit word. A decision error flips the original bit $x_b$ as follows:
\begin{equation}\label{eq:biterror}
\widehat{x}_b = x_b \oplus \epsilon_b
\end{equation}
where $\oplus$ denotes XOR operator and $\epsilon_b=1$ denotes a bit error in $b$th bit position. The decimal representation of the retrieved word is $\widehat{x} = \sum_{b=0}^{B-1}{2^b \widehat{x}_b}$. The decimal error $e$ is given by
\begin{equation}\label{eq:error}
e = \widehat{x} - x = \sum_{B=0}^{B-1}{2^b e_b}
\end{equation} 
where $e_b =\widehat{x}_b - x_b  \in \left\{-1, 0, 1 \right\}$. 

\begin{remark} The decimal error $e = (e_0, \ldots, e_{B-1})$ depends on $x_b$ as well as $\epsilon_b$. Suppose that $\vect{\epsilon} = (1, 0, 0, 1) $. If $x = (1, 0, 0, 1) = 9$, then $\widehat{x} = (0, 0, 0, 0) = 0$, i.e., $e = (-1,0,0,-1) = -9$. If $x = (0, 1, 1, 0) = 6$, then $\widehat{x} = (1, 1, 1, 1) = 15$ and $e = (1,0,0,1) = 9$. 
\end{remark}

Since major noise sources of SRAMs \iffalse such as spatial threshold voltage mismatch and sense amplifier offset \fi are well modeled as Gaussian distributions~\cite{Mizuno1994experimental,Mukhopadhyay2005modeling,Abu-Rahma2012nanometer,Leibowitz2008char}, the error probability of the $b$th bit position is given by
\begin{equation} \label{eq:ber}
p_b = \Pr\left(\epsilon_b = 1\right) = Q\left( \frac{\Delta_b}{\sigma}\right)
\end{equation}
where $\Delta_b$ and $\sigma^2$ denote the swing of $b$th bit position and the noise variance in the corresponding BL, respectively. Note that $Q(x) = \int_x^{\infty}{\frac{1}{\sqrt{2 \pi}} \exp{\left(-\frac{t^2}{2} \right)dt}}$. By increasing $\Delta_b$ in \eqref{eq:ber}, we can reduce $p_b$. However, larger $\Delta_b$ implies more energy consumption and slower speed (see Definitions \ref{def:energy} and \ref{def:max_swing}). 

To measure memory retrieval reliability, bit error probability \eqref{eq:ber} is not appropriate for many applications, since it does not distinguish the differential impact of MSB and LSB errors. Hence, we use the MSE as a fidelity metric. 

\begin{definition}\label{def:mse} The MSE of $x$ is given by
\begin{equation}
\mathsf{MSE}(x) = \mathbb{E}\left[(\widehat{x}-x)^2\right] = \mathbb{E}\left[e^2 \right]. 
\end{equation}
\end{definition}

\begin{lemma}\label{lemma:mse_qfunc}
For a uniformly distributed $x$, $\mathsf{MSE}(x)$ is given by
\begin{equation}\label{eq:mse_qfunc}
\mathsf{MSE}(x) = \mathsf{MSE}(\vect{\Delta}) = \sum_{b=0}^{B-1}{4^b Q\left( \frac{\Delta_b}{\sigma}\right)}. 
\end{equation}	
\end{lemma}
\begin{IEEEproof}
%	See Appendix~\ref{pf:mse_qfucn}. 
	If $x$ is uniformly distributed, the $x_b$s are independent and identically distributed (i.i.d.) and follow the Bernoulli distribution $\mathsf{Ber}\left(\frac{1}{2}\right)$.   
	The MSE of $x$ is given by
	\begin{align}
	\mathsf{MSE}(x) 
	& = \mathbb{E}\left[ \left(\sum_{B=0}^{B-1}{2^b e_b} \right)^2 \right] = \sum_{B=0}^{B-1}{4^b \mathbb{E}\left[e_b^2\right]} 
	= \sum_{b=0}^{B-1}{4^b p_b} \label{eq:pf_mse_1} \\
	&= \sum_{b=0}^{B-1}{4^b Q\left( \frac{\Delta_b}{\sigma}\right)} \label{eq:pf_mse_2}
	\end{align}
%	\begin{align}
%	\mathsf{MSE}(x) 
%	& = \mathbb{E}\left[ \left(\sum_{B=0}^{B-1}{2^b e_b} \right)^2 \right] \label{eq:pf_mse_0} \\
%	&= \sum_{B=0}^{B-1}{4^b \mathbb{E}\left[e_b^2\right]} 
%	= \sum_{b=0}^{B-1}{4^b p_b} \label{eq:pf_mse_1} \\
%	&= \sum_{b=0}^{B-1}{4^b Q\left( \frac{\Delta_b}{\sigma}\right)} \label{eq:pf_mse_2}
%	\end{align}
	where \eqref{eq:pf_mse_1} follows from $\mathbb{E} \left[ e_b^2 \right]= \mathbb{E}\left[ \epsilon_b \right] = p_b$ and $\mathbb{E} \left[ e_i e_j \right] = 0$ since the $e_b$s are independent and $\mathbb{E}\left[e_b \right] = 0$ for $x_b \sim \mathsf{Ber}\left(\frac{1}{2}\right)$~\cite{Yang2011unequal}. In addition, \eqref{eq:pf_mse_2} follows from \eqref{eq:ber}. Because $\mathsf{MSE}(x)$ is a function of $\vect{\Delta}$, we set $\mathsf{MSE}(x) = \mathsf{MSE}(\vect{\Delta})$.  	
\end{IEEEproof}

Note that $\mathsf{MSE}(x)$ is the nonnegative weighted sum of bit error probabilities. The weight $4^b$ represents the differential importance of each bit position. We show that $\mathsf{MSE}(x)$ is convex. 

\begin{lemma}\label{lemma:convex}
	$\mathsf{MSE}(\vect{\Delta})$ is a \emph{convex} function of $\vect{\Delta}$. 
\end{lemma}
\begin{IEEEproof}
	$Q(x)$ is convex for $x \ge 0$ because 
	\begin{equation}
	\frac{d^2Q(x)}{dx^2} = \frac{x}{\sqrt{2\pi}}\exp\left(- \frac{x^2}{2}\right)\ge 0.
	\end{equation}
	Since $\Delta_b \ge 0$ and $\mathsf{MSE}(\vect{\Delta})$ is the nonnegative weighted sum of $Q\left( \frac{\Delta_b}{\sigma}\right)$, $\mathsf{MSE}(\vect{\Delta})$ is convex. 
\end{IEEEproof}
A signed number $x$ can be represented by $x = -x_{B-1} \cdot 2^{B-1} + \sum_{b=0}^{B-2}{2^b x_b}$ whose $\mathsf{MSE}(x)$ is the same as~\eqref{eq:mse_qfunc}. 

%A signed number $x$ can be represented by \begin{equation}\label{eq:signed_representation}
%x = -x_{B-1} \cdot 2^{B-1} + \sum_{b=0}^{B-2}{2^b x_b}
%\end{equation}
%whose $\mathsf{MSE}(x)$ is the same as~\eqref{eq:mse_qfunc}.

Table~\ref{tab:comparison} summarizes the key resource and fidelity metrics for single-bit and $B$-bit word accesses.

\begin{table}[!t]
\renewcommand{\arraystretch}{1.1}
\caption{Resource and Fidelity Metrics for Single-Bit and $B$-bit Word Access}
\vspace{-5mm}
\label{tab:comparison}
\centering
{\hfill{}
	\begin{tabular}{|c|c|c|c|}	\hline
	&  Single bit  & $B$-bit word & Remarks   \\ \hline \hline
	%Variable & $\Delta$  &  $\vect{\Delta} = (\Delta_0, \ldots, \Delta_{B-1})$  \\ \hline
	Energy   & $\Delta$  & $\mathsf{E}(\vect{\Delta}) = \vect{1}^{\mathsf{T}}\vect{\Delta}$ & Definition~\ref{def:energy} \\ \hline
	Delay    & $\Delta$  & $\rho = \max(\vect{\Delta})$ & Definition~\ref{def:max_swing} \\ \hline
	EDP      & $\Delta^2$& $\mathsf{EDP}(\vect{\Delta}) = \mathsf{E}(\vect{\Delta}) \cdot \rho$ & Definition~\ref{def:edp}  \\ \hline \hline
	Fidelity & $p = Q\left(\frac{\Delta}{\sigma}\right)$ & $\mathsf{MSE}(\vect{\Delta}) = \sum{4^b Q\left( \frac{\Delta_b}{\sigma}\right)}$ & \iffalse Definition~\ref{def:mse},\fi Lemma~\ref{lemma:mse_qfunc}  \\ \hline			
	\end{tabular}}
\hfill{}
\vspace{-9mm}
\end{table}

\section{Optimal Bit-Level Swings}\label{sec:proposed}

We formulate convex optimization problems to determine the optimum swings. For a given constraint on MSE, we attempt to (1) minimize energy (low-power SRAMs), (2) maximize speed (high-speed SRAMs), and (3) minimize EDP. Also, we provide generalized water-filling interpretations of these optimization problems based on KKT conditions. 

\subsection{Energy Minimization}

Here, we minimize the read energy for a given constraint on MSE. Hence, we formulate the following convex optimization problem. 

\begin{figure}[!t]
	\centering
	\vspace{-3mm}
	\subfloat[]{\includegraphics[width=0.45\textwidth]{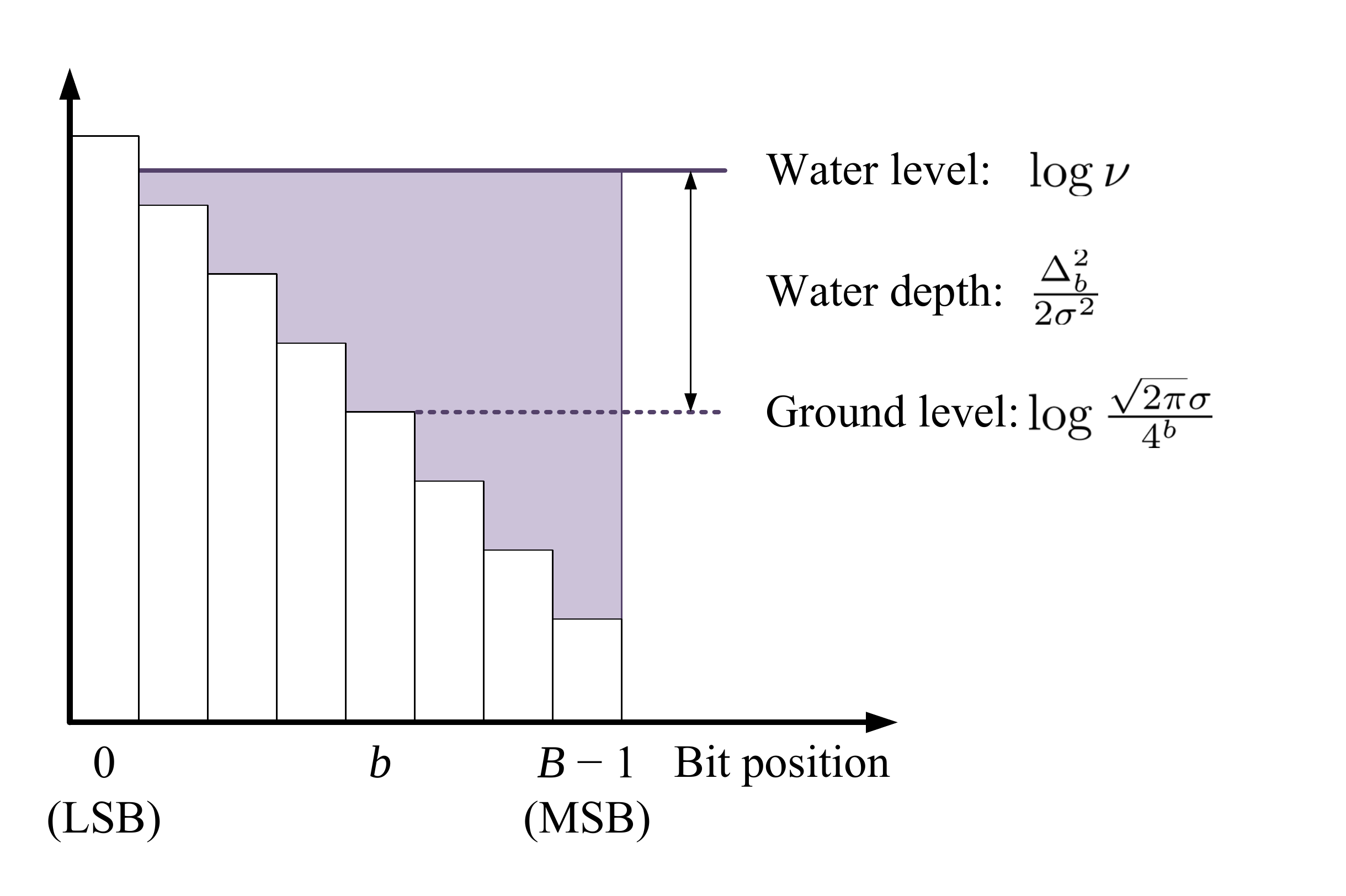}
		\vspace{-10mm}
		\label{fig:min_energy_a}}
	\hfil
	\vspace{-3mm}
	\subfloat[]{\includegraphics[width=0.45\textwidth]{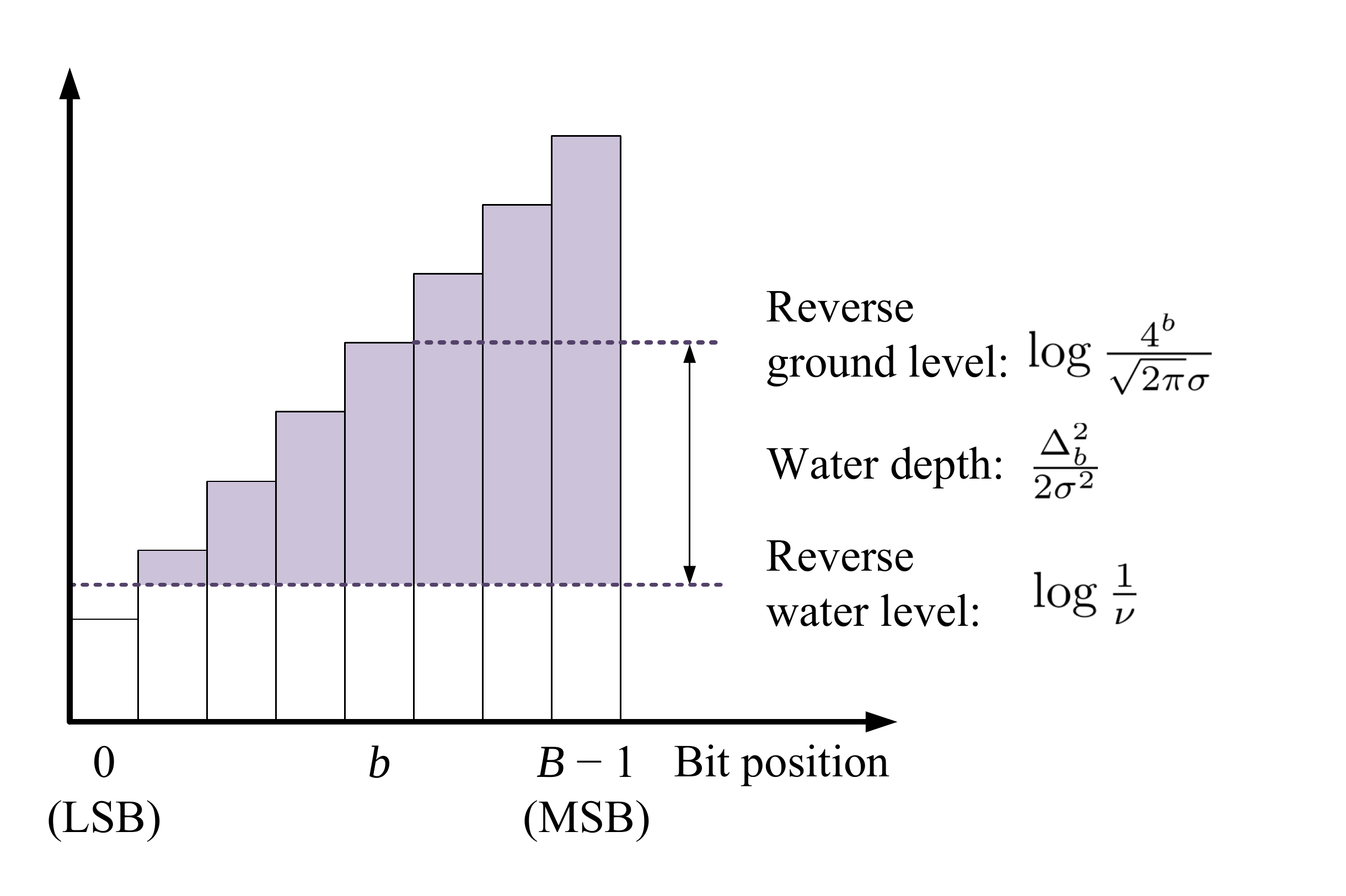}
		\vspace{-10mm}
		\label{fig:min_energy_b}}
	\caption{Graphical interpretations of Theorem~\ref{thm:min_energy}: (a) water-filling and (b) reverse water-filling.}
	\label{fig:min_energy}
	\vspace{-8mm}
\end{figure}	

\begin{equation}
\begin{aligned} \label{eq:min_energy}
& \underset{\vect{\Delta}}{\text{minimize}}
& & \mathsf{E}(\vect{\Delta}) = \vect{1}^{\mathsf{T}} \vect{\Delta} \\
&{\text{subject~to}} & & \sum_{b=0}^{B-1}{4^b Q\left( \frac{\Delta_b}{\sigma}\right)} \le \mathcal{V} \\
& & & \Delta_b \ge 0, \quad b=0,\ldots,B-1
\end{aligned}
\end{equation}
where $\mathcal{V}$ is a constant corresponding to the given constraint of MSE. 

Since the objective and constraints are convex, the optimization problem \eqref{eq:min_energy} is convex. The optimal solution can be derived by KKT conditions. 
\begin{theorem}\label{thm:min_energy}The optimal swing $\vect{\Delta}^*$ of \eqref{eq:min_energy} is given by
	\begin{equation}\label{eq:min_energy_sol}
	\Delta_b^* =
	\begin{cases}
	0, & \text{if}\: \nu \le \frac{\sqrt{2\pi}\sigma}{4^b}, \\
	\sigma \sqrt{2 \log{\left(\frac{4^b}{\sqrt{2\pi}\sigma} \cdot \nu \right)}}, & \text{otherwise}
	\end{cases}
	\end{equation}
	where $\nu$ is a dual variable. 	
\end{theorem}
\begin{IEEEproof}
	We define the Lagrangian $L_1(\vect{\Delta}, \nu, \vect{\lambda})$ associated with problem \eqref{eq:min_energy} as
	\begin{align}
	L_1(\vect{\Delta}, \nu, \vect{\lambda}) = \vect{1}^{\mathsf{T}} \vect{\Delta} + \nu \left( \sum_{b=0}^{B-1}{4^b Q\left( \frac{\Delta_b}{\sigma}\right)} - \mathcal{V} \right) - \sum_{b=0}^{B-1}{\lambda_b \Delta_b} 	
	\end{align}
	where $\nu$ and $\vect{\lambda}=(\lambda_0,\ldots,\lambda_{B-1})$ are the dual variables. The optimal solution \eqref{eq:min_energy_sol} is derived from $L_1$ and the KKT conditions. The details of the proof are given in Appendix~\ref{pf:min_energy}.
\end{IEEEproof}

%\begin{remark}[Water-filling]
The optimal solution \eqref{eq:min_energy_sol} can be interpreted as classical \emph{water-filling} or \emph{reverse water-filling} as shown in Fig.~\ref{fig:min_energy}. Each bit position can be regarded as an individual channel among $B$ parallel channels. In the water-filling interpretation (see Fig.~\ref{fig:min_energy}\subref{fig:min_energy_a}), the ground levels depend on the importance of bit positions. We flood the bins to the water level of $\log{\nu}$. Since the MSB has the lowest ground level and the LSB has the highest ground level, larger swings are assigned to more significant bit positions. For a bit position $b$ such that $\nu > \frac{\sqrt{2\pi}\sigma}{4^b}$, we can readily obtain the following equation (see Appendix~\ref{pf:min_energy}):
\begin{equation} \label{eq:min_energy_relation}
\log{\nu} = \log{\frac{\sqrt{2\pi}\sigma}{4^b}} + \frac{\Delta_b^2}{2 \sigma^2}
\end{equation} 
where $\log{\nu}$, $\log{\frac{\sqrt{2\pi}\sigma}{4^b}}$, and $\frac{\Delta_b^2}{2 \sigma^2}$ represent the water level, the ground level, and the water depth, respectively. The water level $\log{\nu}$ depends on $\mathcal{V}$ in \eqref{eq:min_energy}. 

Fig.~\ref{fig:min_energy}\subref{fig:min_energy_b} illustrates a reverse water-filling interpretation of \eqref{eq:min_energy_sol}. For a bit position $b$ such that $\frac{1}{\nu} < \frac{4^b}{\sqrt{2\pi}\sigma}$, by modifying \eqref{eq:min_energy_relation}, we can readily obtain
\begin{equation} \label{eq:min_energy_relation_reverse}
\log{\frac{4^b}{\sqrt{2\pi}\sigma}} = \log{\frac{1}{\nu}} + \frac{\Delta_b^2}{2 \sigma^2}
\end{equation} 
where $\log{\frac{4^b}{\sqrt{2\pi}\sigma}}$ and $\log{\frac{1}{\nu}}$ denote the reverse ground level and the reverse water level, respectively. The reverse ground level implies the importance of each bit position. We allocate positive swings only for bit positions whose reverse ground levels are greater than the reverse water level. 

Although we are dealing with the weighted bit error probabilities $4^b Q\left(\frac{\Delta_b}{\sigma}\right)$  rather than capacities (for water-filling) or rate distortion functions (for reverse water-filling), we still obtain water-filling and reverse water-filling interpretations. 

\begin{remark}[LSB dropping and constant-power water-filling]\label{rem:LSB_dropping} Constant-power water-filling activates the subset of parallel channels but with a constant power allocation~\cite{Chow1993bandwidth,Yu2001constant}. Constant-power water-filling in communication theory is equivalent to LSB dropping in circuit theory~\cite{Frustaci2015sram,Frustaci2016approximate,Kaul2012variable} since LSB dropping allocates uniform swings for undropped bit positions. 
\end{remark}

\subsection{Speed Maximization}

Here, we maximize the speed of read access for a given constraint on MSE. The maximum speed can be achieved by minimizing $\rho$ of \eqref{eq:max_swing} since $\rho$ is proportional to the maximum pulse-width $T_{\text{max}}$.  

\begin{equation}
\begin{aligned} \label{eq:max_speed0}
& \underset{\vect{\Delta}}{\text{minimize}}
& & \rho = \max\left\{\Delta_0, \ldots, \Delta_{B-1}\right\} \\
&{\text{subject~to}} & & \sum_{b=0}^{B-1}{4^b Q\left( \frac{\Delta_b}{\sigma}\right)} \le \mathcal{V} \\
& & & \Delta_b \ge 0, \quad b=0,\ldots,B-1
\end{aligned}
\end{equation}
By introducing an additional variable $\xi$, we can reformulate \eqref{eq:max_speed0} as
\begin{equation}
\begin{aligned} \label{eq:max_speed}
& \underset{\vect{\Delta}, \xi}{\text{minimize}}
& & \xi \\
&{\text{subject~to}} & & \sum_{b=0}^{B-1}{4^b Q\left( \frac{\Delta_b}{\sigma}\right)} \le \mathcal{V} \\
& & & 0 \le \Delta_b \le \xi, \quad b=0,\ldots,B-1 
%& & & \Delta_b \le \xi, \quad b=0,\ldots,B-1
\end{aligned}
\end{equation}
This reformulated optimization problem is also convex. From KKT conditions, we show that $\xi = \rho$ (see Appendix~\ref{pf:max_speed}). 

\begin{theorem}\label{thm:max_speed}The optimal swing $\vect{\Delta}^*$ of \eqref{eq:max_speed0} is given by
	\begin{equation}\label{eq:max_speed_sol}
	\Delta_b^* = \rho = \xi = \sigma \sqrt{2 \log{\left(\frac{4^B-1}{3\sqrt{2\pi}\sigma}\cdot \nu\right)}}	
	\end{equation}
	for all $b\in[0, B-1]$. Note that $\nu$ is a dual variable.  
\end{theorem}	
\begin{IEEEproof}
	We define the Lagrangian $L_2(\vect{\Delta}, \xi, \nu, \vect{\lambda}, \vect{\eta})$ associated with problem \eqref{eq:max_speed} as
	\begin{align}
	L_2 (\vect{\Delta}, \xi, \nu, \vect{\lambda}, \vect{\eta}) = \xi + \nu \left( \sum_{b=0}^{B-1}{4^b Q\left( \frac{\Delta_b}{\sigma}\right)} - \mathcal{V} \right) - \sum_{b=0}^{B-1}{\lambda_b \Delta_b} 
	+ \sum_{b=0}^{B-1}{\eta_b (\Delta_b - \xi)}
	\end{align}
	where $\nu$, $\vect{\lambda}=(\lambda_0,\ldots,\lambda_{B-1})$ and $\vect{\eta}=(\eta_0,\ldots,\eta_{B-1})$ are dual variables. The optimal solution  \eqref{eq:max_speed_sol} can be derived from $L_2$ and corresponding KKT conditions. The details of the proof are given in Appendix~\ref{pf:max_speed}.
\end{IEEEproof}

%\begin{remark}[Ground-flattening and Water-filling] 
%\end{remark}

\begin{figure}[!t]
	\centering
	\vspace{-8mm}
	\subfloat[]{\includegraphics[width=0.45\textwidth]{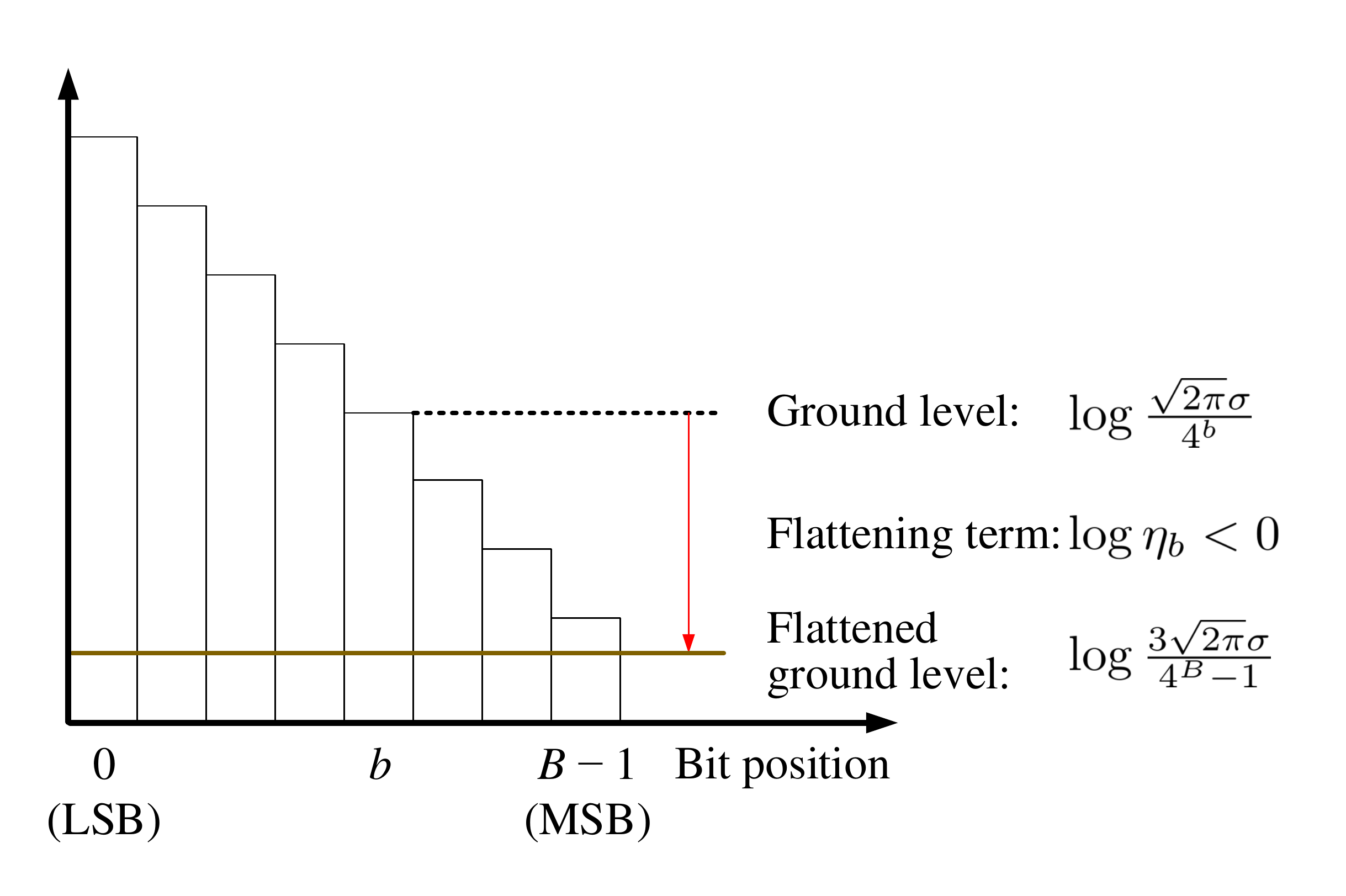}
		\label{fig:max_speed_a}}
	\hfil
	\vspace{-2mm}
	\subfloat[]{\includegraphics[width=0.45\textwidth]{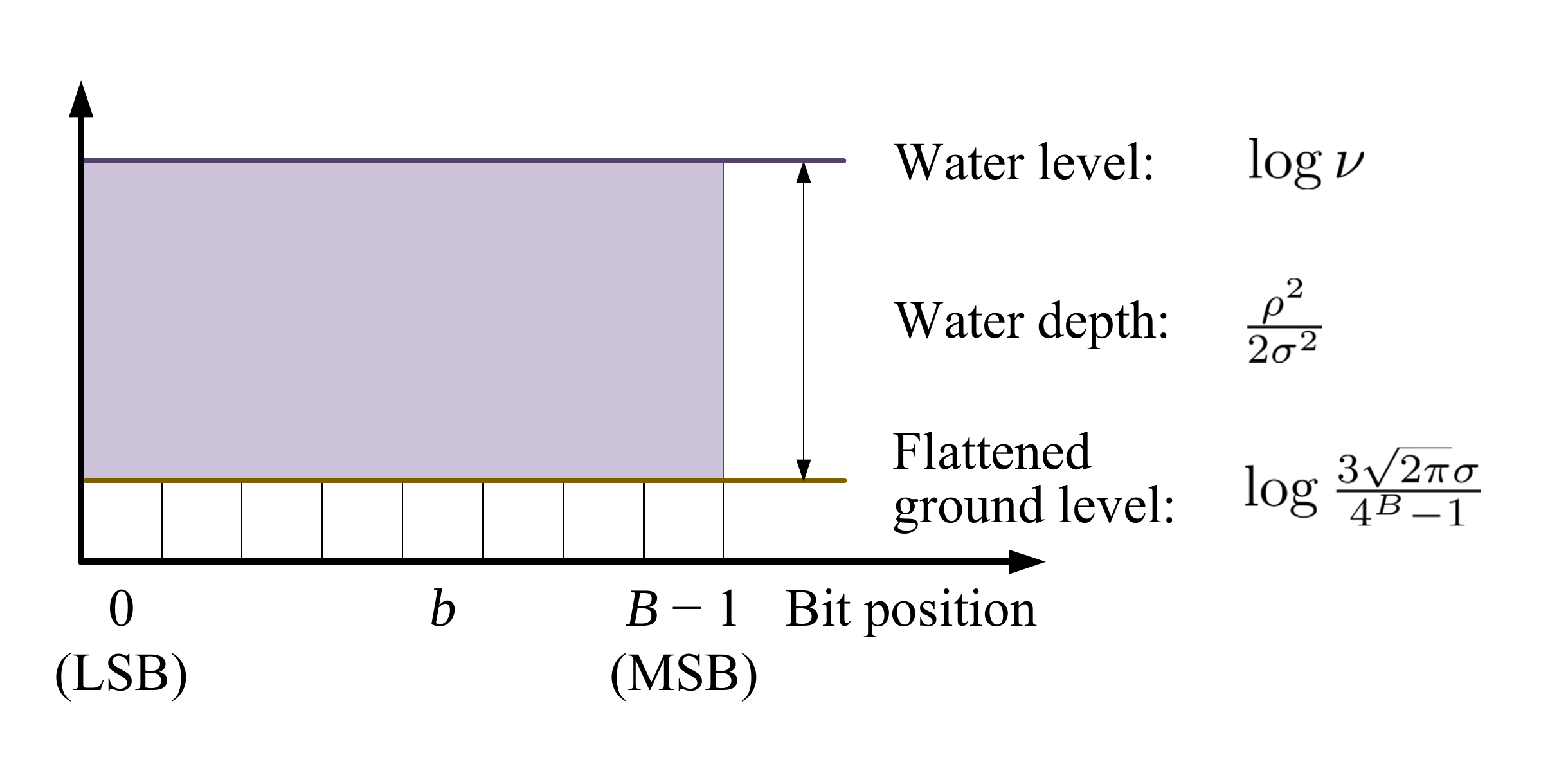}
		\label{fig:max_speed_b}}
	\caption{Ground-flattening and water-filling interpretation of Theorem~\ref{thm:max_speed}: (a) ground-flattening and (b) water-filling (after ground-flattening).}
	\label{fig:max_speed}
	\vspace{-8mm}
\end{figure}	

The optimal solution \eqref{eq:max_speed_sol} can be interpreted as \emph{ground-flattening} and \emph{water-filling}. For any $b \in [0, B-1]$, we derive the following equation (see Appendix~\ref{pf:max_speed}): 
	\begin{equation} \label{eq:max_speed_relation}
	\log{\nu} = \log{\frac{\sqrt{2\pi}\sigma}{4^b}} + \log{\eta_b} + \frac{\Delta_b^2}{2 \sigma^2}
	\end{equation}	
where $\log{\nu}$, $\log{\frac{\sqrt{2\pi}\sigma}{4^b}}$, $\log{\eta_b}$, and $\frac{\Delta_b^2}{2 \sigma^2}$ represent the water level, the ground level, the ground-flattening term, and the water depth, respectively. Compared with \eqref{eq:min_energy_relation}, we observe that \eqref{eq:max_speed_relation} has an additional ground-flattening term $\log{\eta_b}$. 
By solving KKT conditions, we show that 
\begin{equation}\label{eq:max_speed_eta}
\log \eta_b = \log \frac{3}{4^B-1} \cdot 4^b < 0. 
\end{equation}
Hence, the \emph{flattened ground level} (i.e., the sum of the ground level and the ground flattening term) is given by
\begin{equation} \label{eq:max_speed_flattened_ground}
\log{\frac{\sqrt{2\pi}\sigma}{4^b}} + \log{\eta_b} = \log{\frac{3\sqrt{2\pi}\sigma}{4^B-1}}.
\end{equation}
Since the unequal ground levels are flattened by the flattening terms, the water depths of all bit positions are identical after water-filling (see Fig.~\ref{fig:max_speed}\subref{fig:max_speed_b}). 

\begin{figure}[t]
	\centering
	\vspace{-8mm}
	\subfloat[]{\includegraphics[width=0.45\textwidth]{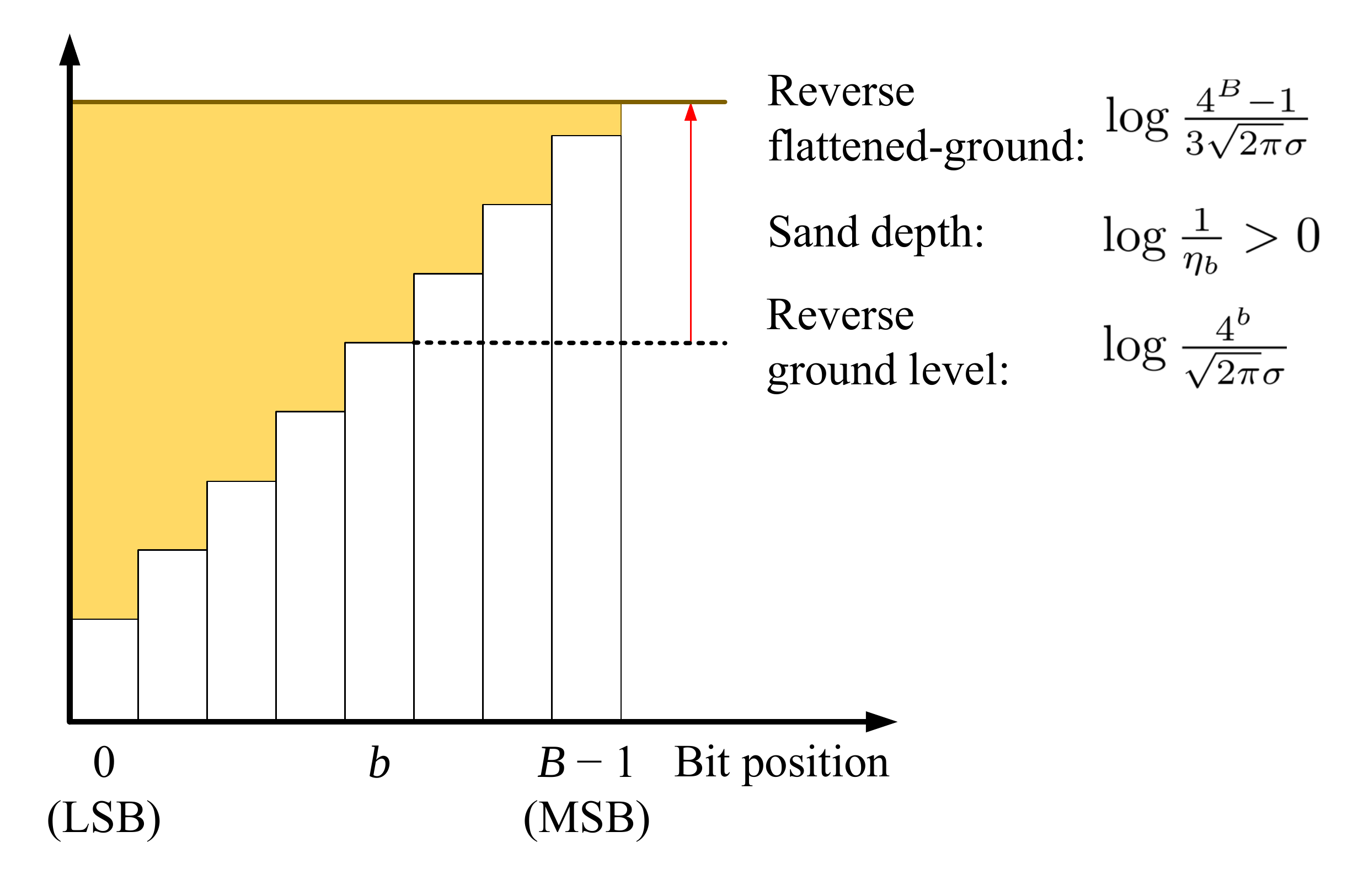}
	\label{fig:max_speed_a_reverse}}
	\hfil
	\vspace{-3mm}
	\subfloat[]{\includegraphics[width=0.45\textwidth]{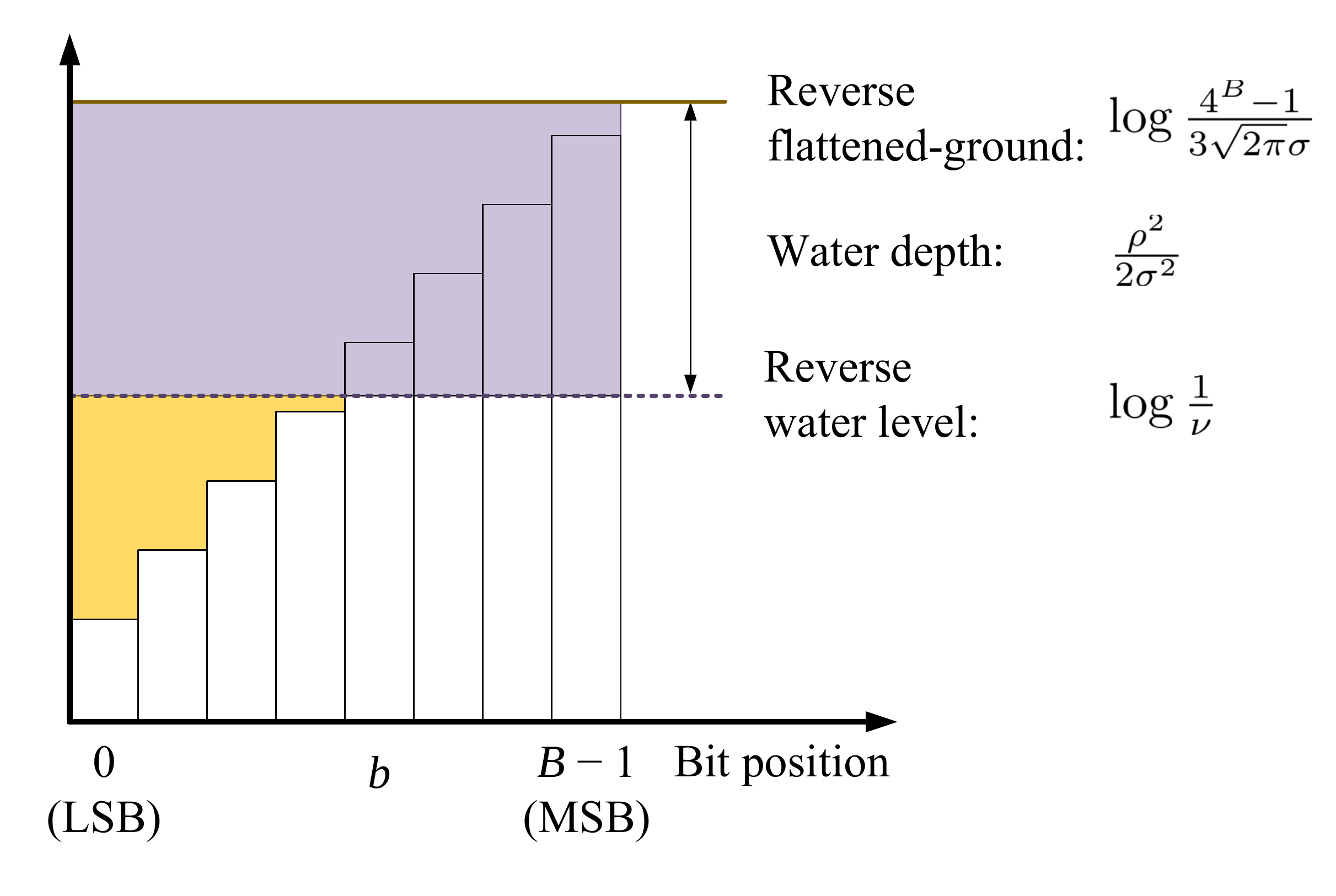}
		\label{fig:max_speed_b_reverse}}
	\caption{Sand-pouring and reverse water-filling interpretation of Theorem~\ref{thm:max_speed}: (a) sand-pouring and (b) reverse water-filling (after sand-pouring).}
	\label{fig:max_speed_reverse}
	\vspace{-6mm}
\end{figure}	

In addition, the optimal solution \eqref{eq:max_speed_sol} can be interpreted by \emph{sand-pouring} and \emph{reverse water-filling}. We can modify \eqref{eq:max_speed_flattened_ground} into
\begin{equation} \label{eq:max_speed_flattened_ground_reverse}
\log{\frac{4^b}{\sqrt{2\pi}\sigma}} + \log{\frac{1}{\eta_b}} = \log{\frac{4^B-1}{3\sqrt{2\pi}\sigma}}.
\end{equation}	
where $\log{\frac{4^b}{\sqrt{2\pi}\sigma}}$, $\log\frac{1}{\eta_b}$, and $\log{\frac{4^B-1}{3\sqrt{2\pi}\sigma}}$ represent the reverse ground level, the sand depth, and the reverse flattened ground level, respectively. The positive sand depth (see \eqref{eq:max_speed_eta}) fills the gap between each reverse ground level and the reverse flattened ground level (see Fig.~\ref{fig:max_speed_reverse}\subref{fig:max_speed_a_reverse}). The reverse flattened ground results in uniform swings as shown in Fig.~\ref{fig:max_speed_reverse}\subref{fig:max_speed_b_reverse}. 

\begin{remark} \label{remark:max_speed}
	Conventional uniform swing assignment maximizes the read access speed if importance of bit positions is ignored. 
\end{remark}

\begin{remark} \label{remark:max_speed_mse}
	For conventional uniform swings assignment, the MSE is given by 
	\begin{equation}
	\mathsf{MSE}(x) = \frac{4^B-1}{3} \cdot p
	\end{equation}
	which comes from Lemma~\ref{lemma:mse_qfunc} and $p_b = p$ for any $b\in\{0, \ldots, B-1\}$. 	
\end{remark}

\begin{remark} \label{remark:min_ber}
	The overall bit error rate (BER) is the sum of bit error probabilities of all bit positions as follows:
	\begin{equation}
	\mathsf{BER}=\sum_{b=0}^{B-1}{Q \left(\frac{\Delta_b}{\sigma}\right)}
	\end{equation} 
Since $Q(\cdot)$ is convex (see the proof of Lemma~\ref{lemma:convex}), the uniform swing assignment minimizes the overall BER. 	
\end{remark}

If we do not consider the differential importance of each bit position, the conventional uniform swing is optimal since it maximizes the read access speed (Remark~\ref{remark:max_speed}) and minimizes the overall BER (Remark~\ref{remark:min_ber}). 

\subsection{EDP Minimization}

We formulate the following convex optimization problem to minimize EDP for a given constraint on MSE. 

\begin{equation}
\begin{aligned} \label{eq:min_edp}
& \underset{\vect{\Delta}, \xi}{\text{minimize}}
& & \vect{1}^{\mathsf{T}} \vect{\Delta} \cdot \xi \\
&{\text{subject~to}} & & \sum_{b=0}^{B-1}{4^b Q\left( \frac{\Delta_b}{\sigma}\right)} \le \mathcal{V} \\
& & & 0 \le \Delta_b \le \xi, \quad b=0,\ldots,B-1 
%& & & \Delta_b \le \xi, \quad b=0,\ldots,B-1
\end{aligned}
\end{equation}
which is derived by taking into account~\eqref{eq:edp} and~\eqref{eq:max_speed}. We show that $\xi$ is equal to $\rho$ (see Appendix~\ref{pf:min_edp}). %Note that the optimal solution can be obtained by standard algorithms since \eqref{eq:min_edp} is a convex optimization problem. 

\begin{theorem}\label{thm:min_edp}The optimal swing $\vect{\Delta}^*$ of \eqref{eq:min_edp} is given by
\begin{align}\label{eq:min_edp_sol}
\Delta_b^* = 
\begin{cases}
0, & \text{if}\: \log{\frac{\nu}{\rho}} \le \log{\frac{\sqrt{2\pi}\sigma}{4^b}}, \\
\rho, & \text{if}\: \log{\frac{\nu}{\rho}} \ge \log{\frac{\sqrt{2\pi}\sigma}{4^b}} + \frac{\rho^2}{2 \sigma^2}, \\
\sigma \sqrt{2 \log{\left(\frac{4^b}{\sqrt{2\pi}\sigma} \cdot \frac{\nu}{\rho}\right)}}, &
\text{otherwise}
%\text{if}\: \log{\frac{\sqrt{2\pi}\sigma}{4^b}} \le \log{\frac{\nu}{\rho}} \le \log{\frac{\sqrt{2\pi}\sigma}{4^b}} + \frac{\rho^2}{2 \sigma^2}
\end{cases}
\end{align}
%\begin{numcases}{\Delta_b^*=\hspace{-5mm}}
%0, & \hspace{-7mm} if $\log{\frac{\nu}{\rho}} \le \log{\frac{\sqrt{2\pi}\sigma}{4^b}}$, 
%\\
%\rho, & \hspace{-7mm} if $\log{\frac{\nu}{\rho}} \ge \log{\frac{\sqrt{2\pi}\sigma}{4^b}} + \frac{\rho^2}{2 \sigma^2}$, 
%\\
%\sigma \sqrt{2 \log{\left(\frac{4^b}{\sqrt{2\pi}\sigma} \cdot \frac{\nu}{\rho}\right)}}, & \hspace{-7mm} otherwise
%\end{numcases}
where $\nu$ is a dual variable.  
\end{theorem}	
\begin{IEEEproof}
	We define the Lagrangian $L_3(\vect{\Delta}, \xi, \nu, \vect{\lambda}, \vect{\eta})$ associated with problem \eqref{eq:min_edp} as
	\begin{align}
	L_3 (\vect{\Delta}, \xi, \nu, \vect{\lambda}, \vect{\eta}) 
	= \vect{1}^{\mathsf{T}}\vect{\Delta}\cdot\xi + \nu \left( \sum_{b=0}^{B-1}{4^b Q\left( \frac{\Delta_b}{\sigma}\right)} - \mathcal{V} \right) - \sum_{b=0}^{B-1}{\lambda_b \Delta_b}  
	+ \sum_{b=0}^{B-1}{\eta_b (\Delta_b - \xi)}
	\end{align}
	where $\nu$, $\vect{\lambda}=(\lambda_0,\ldots,\lambda_{B-1})$, and $\vect{\eta}=(\eta_0,\ldots,\eta_{B-1})$ are dual variables. The optimal solution \eqref{eq:min_edp_sol} can be derived from $L_3$ and corresponding KKT conditions. The details of the proof are given in Appendix~\ref{pf:min_edp}.
\end{IEEEproof}

%\begin{equation}
%\Delta_b^* = \left\{ \,
%\begin{IEEEeqnarraybox}[][c]{l?s}
%\IEEEstrut
%0, & if $\log{\frac{\nu}{\rho}} \le \log{\frac{\sqrt{2\pi}\sigma}{4^b}}$, \\
%\rho, & if $ \log{\frac{\nu}{\rho}} \ge \log{\frac{\sqrt{2\pi}\sigma}{4^b}} + \frac{\rho^2}{2 \sigma^2}$, \\
%\sigma \sqrt{2 \log{\left(\frac{4^b}{\sqrt{2\pi}\sigma} \cdot \frac{\nu}{\rho}\right)}}, & otherwise.
%\IEEEstrut
%\end{IEEEeqnarraybox}
%\right
%\label{eq:example_left_right1}
%\end{equation}.

\begin{figure}[t]
	\centering
	\vspace{-4mm}
	\subfloat[]{\includegraphics[width=0.45\textwidth]{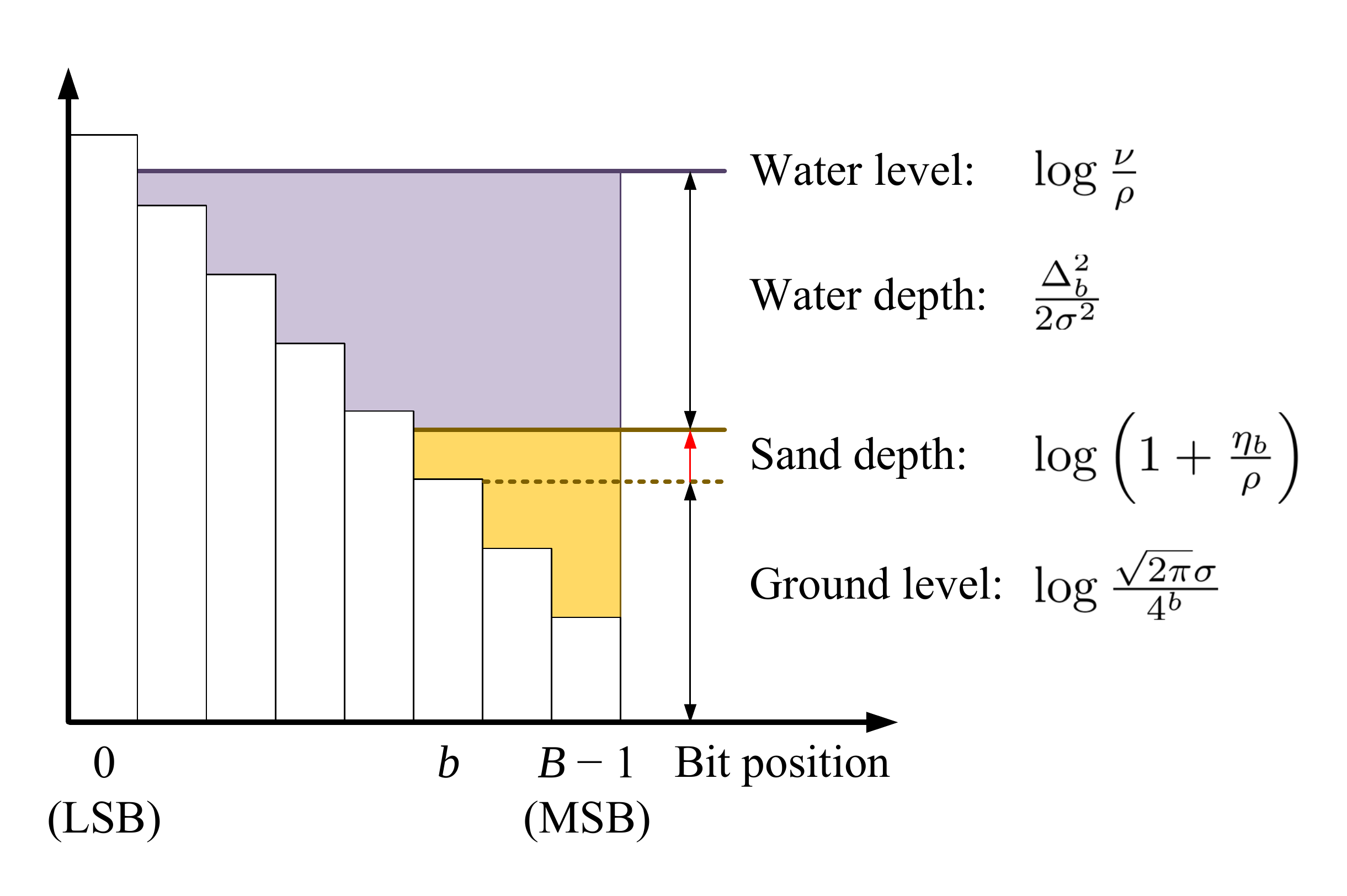}		\label{fig:min_edp_a}}
	\hfil
	\subfloat[]{\includegraphics[width=0.45\textwidth]{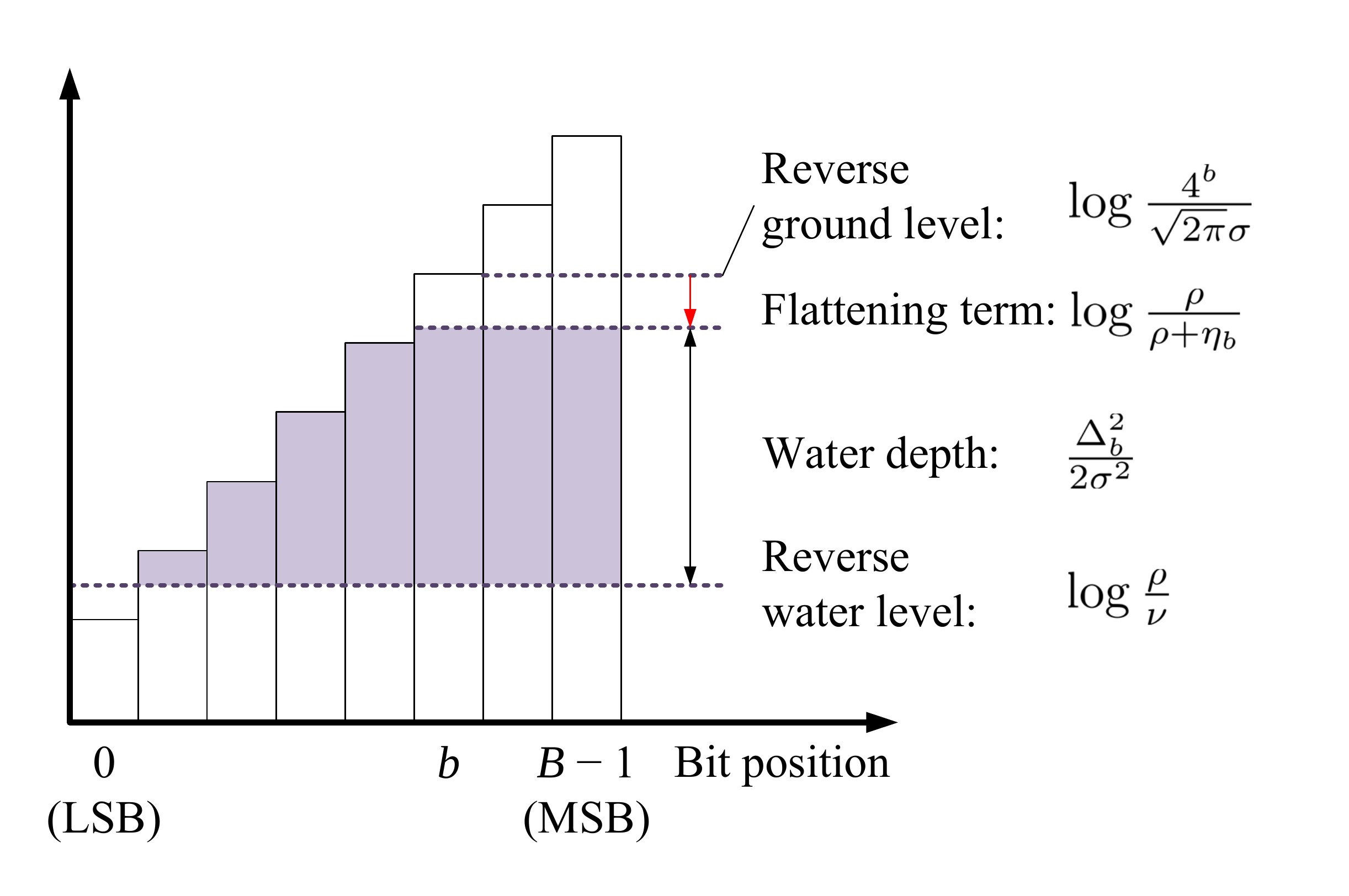}
		\label{fig:min_edp_b}}
	\vspace{-3mm}	
	\caption{Graphical interpretations of Theorem~\ref{thm:min_edp}: (a) sand-pouring and water-filling and (b) ground-flattening and reverse water-filling.}
	\label{fig:min_edp}
	\vspace{-8mm}
\end{figure}

The optimal solution of \eqref{eq:min_edp_sol} can be interpreted by \emph{sand-pouring} and \emph{water-filling} as shown in Fig.~\ref{fig:min_edp}\subref{fig:min_edp_a}. For $\log{\frac{\nu}{\rho}} > \log{\frac{\sqrt{2\pi}\sigma}{4^b}}$, we derive the following equation (see Appendix~\ref{pf:min_edp}): 
\begin{equation}\label{eq:min_edp_relation}
\log{\frac{\nu}{\rho}} = \log{\frac{\sqrt{2\pi}\sigma}{4^b}} + \log{\left(1 + \frac{\eta_b}{\rho} \right)} + \frac{\Delta_b^2}{2 \sigma^2}
\end{equation}
where $\log{\frac{\nu}{\rho}}$, $\log{\frac{\sqrt{2\pi}\sigma}{4^b}}$, $\log{\left(1 + \frac{\eta_b}{\rho} \right)}$, and $\frac{\Delta_b^2}{2 \sigma^2}$ represent the water level, the ground level, the sand depth, and the water depth, respectively. \emph{Pouring sand} suppresses the maximum water depth (i.e., the maximum swing) and \emph{water-filling} allocates swings by taking into account energy efficiency. 

The following corollary shows the relation between the sand depth and other metrics.
\begin{corollary}\label{cor:sand_depth}
The sand depth $s_b$ is given by
%\begin{align}\label{eq:sand_depth}
%s_b = 
%\begin{cases}
%0, & \text{if}\: 0 \le \Delta_b < \rho, \\
%\log{\left(1 + \frac{\eta_b}{\rho} \right)}, & \text{if}\: \Delta_b = \rho. 
%\end{cases}
%\end{align}
\begin{equation}\label{eq:sand_depth}
s_b = \log{\left(1 + \frac{\eta_b}{\rho} \right)} 
\end{equation}
where
\begin{equation}\label{eq:sand_eta}
\eta_b = \begin{cases}
0, & \text{if}~0 \le \Delta_b < \rho, \\
>0, & \text{if}~\Delta_b = \rho.
\end{cases}
\end{equation}
Hence, $s_b = 0$ for $0 \le \Delta_b < \rho$ and $s_b > 0$ for $\Delta_b = \rho$. Also, the amount of sand is given by
\begin{equation}\label{eq:sand_depth_amount}
\sum_{b=0}^{B-1}{\exp(s_b)} = \frac{\mathsf{E}(\vect{\Delta})}{\rho} + B. 
\end{equation}
\begin{IEEEproof}
	See Appendix~\ref{pf:min_edp}.
\end{IEEEproof}	
%\begin{IEEEproof}
%	From KKT conditions, we obtain $\sum_{b=0}^{B-1}{\Delta_b} = \sum_{b=0}^{B-1}{\eta_b}$ (see \eqref{eq:cr3_KKT_gr_2} in Appendix~\ref{pf:min_edp}). By combining this condition and \eqref{eq:sand_depth}, \eqref{eq:sand_depth_amount} can be derived. 
%\end{IEEEproof}	
\end{corollary}
We observe that the amount of sand depends on the energy and the maximum swing. \iffalse More energy (i.e., water) increases the amount of sand. On the other hand, larger maximum swing (i.e., the maximum water depth) implies less sand. \fi 

Suppose that sand is poured in only the MSB position, i.e., $\Delta_{B-1} = \rho$ and $\Delta_{b} < \rho$ for $b \in [0, B-2]$. Then,
\begin{align}
\eta_{B-1} &= \sum_{b=0}^{B-1}{\eta_b} = \sum_{b=0}^{B-1}{\Delta_b} = \mathsf{E}(\vect{\Delta})
\end{align}
which follows from \eqref{eq:sand_eta}, \eqref{eq:cr3_KKT_gr_2} (in Appendix~\ref{pf:min_edp}), and Definition~\ref{def:energy}. Hence,
\begin{equation}\label{eq:sand_capacity}
s_{B-1} = \log{\left(1 + \frac{\mathsf{E}(\vect{\Delta})}{\rho}\right)} = \log{\left(1 + \frac{B}{\mathsf{PASR}(\vect{\Delta})}\right)}
\end{equation}
where the peak-to-average swing ratio (PASR) of swings is given by
\begin{equation}\label{eq:pasr}
\mathsf{PASR}(\vect{\Delta}) = \frac{\rho}{\frac{1}{B}\cdot\mathsf{E}{(\vect{\Delta}})}. 
\end{equation}
We also note that \eqref{eq:sand_capacity} takes a similar form as the Gaussian channel's capacity. By \eqref{eq:sand_capacity} and \eqref{eq:pasr}, we obtain
\begin{equation}
\mathsf{PASR}(\vect{\Delta}) = \frac{B}{\exp{(s_{B-1})}-1}
\end{equation}  
which shows that more sand reduces the PASR of swings. 

Fig.~\ref{fig:min_edp}\subref{fig:min_edp_b} illustrates the \emph{ground-flattening} and \emph{reverse water-filling} interpretation. From~\eqref{eq:min_edp_relation}, we can obtain
\begin{equation}\label{eq:min_edp_relation_reverse}
\log{\frac{4^b}{\sqrt{2\pi}\sigma}} + \log{\frac{\rho}{\rho+\eta_b}} = \log{\frac{\rho}{\nu}} + \frac{\Delta_b^2}{2 \sigma^2}
\end{equation}
where the negative flattening term $\log{\frac{\rho}{\rho+\eta_b}}$ suppresses the maximum swing and reverse water-filling up to the reverse water level $\log\frac{\rho}{\nu}$ optimizes energy efficiency. 

%\begin{table}[!t]
%	\renewcommand{\arraystretch}{1.1}
%	\caption{Summary of Generalized Water-filling}
%	\vspace{-4mm}
%	\label{tab:waterfilling}
%	\centering
%	{\hfill{}
%		\begin{tabular}{|c|c|c|c|}	\hline
%			&  Water-filling  & Reverse water-filling & Ground  \\
%			&  interpretation & interpretation        & levels  \\ \hline \hline
%			Min energy   & Water-filling  & Reverse water-filling & Unflattened \\ \hline
%			\multirow{2}{*}{Max speed} & Ground-flattening/  & Sand-pouring/ & Perfectly  \\
%			& water-filling & reverse water-filling & flattened \\ \hline
%			\multirow{2}{*}{Min EDP} & Sand-pouring/ & Ground-flattening/ & Partially \\
%			& water-filling & reverse water-filling & flattened \\ \hline
%	\end{tabular}}
%	\hfill{}
%	\vspace{-8mm}
%\end{table} 

\begin{table}[!t]
	\renewcommand{\arraystretch}{1.1}
	\caption{Summary of Generalized Water-filling}
	\vspace{-4mm}
	\label{tab:waterfilling}
	\centering
	{\hfill{}
		\begin{tabular}{|c|c|c|c|}	\hline
			&  Water-filling interpretation & Reverse water-filling interpretation & Ground levels \\ \hline \hline
			Min energy   & Water-filling  & Reverse water-filling & Unflattened \\ \hline
			{Max speed} & Ground-flattening / water-filling  & Sand-pouring / reverse water-filling & Perfectly flattened \\ \hline
			{Min EDP} & Sand-pouring / water-filling & Ground-flattening / reverse water-filling & Partially flattened \\ \hline
	\end{tabular}}
	\hfill{}
	\vspace{-8mm}
\end{table}

\begin{remark}[Sand-pouring and mercury-filling] \emph{Sand-pouring} and \emph{water-filling} has a connection to mercury/water-filling \cite{Lozano2006optimum} because both are explained by two-level filling. In the mercury/water-filling problem, the mercury is poured before water-filling to fill the gap between an ideal Gaussian signal and practical signal constellations, hence, each mercury depth depends only on the corresponding signal constellation. On the other hand, sand-pouring depends on the ground level and sand depths are correlated with each other since sand-pouring attempts to flatten the ground. Also, the amount of poured sand depends on water-filling as shown in Corollary~\ref{cor:sand_depth} whereas the amount of mercury is not related to water-filling. \end{remark}

\begin{remark}[Ground-flattening and Sand-pouring]
	The terms \emph{ground-flattening} and \emph{sand-pouring} come from analogies with hydrodynamics. In hydrodynamics, flattening ground levels increases the flow speed by reducing \emph{wetted perimeter}\footnote{The wetted perimeter is the perimeter of the cross-sectional area that is in contact with the aqueous body. Friction losses typically increase with an increasing wetted perimeter.}~\cite{Knighton1998fluvial}. In our optimization problems, ground-flattening terms in \eqref{eq:max_speed_relation} maximize the read speed by achieving perfectly even ground levels. The sand-pouring of \eqref{eq:min_edp_relation} limits the speed performance degradation by partially flattening the ground levels.  
\end{remark}

Table~\ref{tab:waterfilling} summarizes water-filling and reverse water-filling interpretations for our optimization problems. Notice the duality between ground-flattening and sand-pouring. 

\section{Non-uniform Sources and Non-Gaussian Noises}\label{sec:extension}

In this section, we study how to extend our optimization problems to non-uniformly distributed sources and to non-Gaussian noise models. 

\subsection{Non-uniform Sources}\label{subsec:nonuniform}

In~Lemma~\ref{lemma:mse_qfunc}, we considered the MSE of a uniformly distributed source. For a non-uniformly distributed source $x = \sum_{b=0}^{B-1}{x_b}$ of \eqref{eq:representation}, the MSE is derived in the following proposition. 

\begin{proposition}The MSE of $x$ is given by
	\begin{align} \label{eq:mse_any}
	\mathsf{MSE}(x) = \sum_{b=0}^{B-1}{4^b p_b } + 2 \sum_{b=1}^{B-1}{\sum_{b'=0}^{b-1}{2^{b+b'}p_b p_{b'} \phi(b, b')}} 
	\end{align}
	where $\phi(b,b')= \Pr\left(x_b=x_{b'}\right) - \Pr\left(x_b\ne x_{b'}\right)$, $p_b = Q\left(\frac{\Delta_b}{\sigma}\right)$, and $p_b' = Q\left(\frac{\Delta_{b'}}{\sigma}\right)$. 
\end{proposition}
\begin{IEEEproof}
	From \eqref{eq:pf_mse_1}, the MSE of $x$ is given by 
	\begin{align}
	\mathsf{MSE}(x) 
	& = \mathbb{E}\left[ \left(\sum_{B=0}^{B-1}{2^b e_b} \right)^2 \right] = \sum_{B=0}^{B-1}{4^b p_b} + 2 \sum_{b=1}^{B-1}{\sum_{b'=0}^{b-1}{2^{b+b'}\mathbb{E}\left[e_b e_{b'} \right]}}     
	\end{align}
	where $\mathbb{E}\left[e_b e_{b'} \right]$ for $b\ne b'$ is given by
	\begin{align}
	\mathbb{E}\left[e_b e_{b'} \right] &= \sum_{x, \widehat{x}}{p(x) p(\widehat{x} \mid x)e_b e_{b'}} =p_b p_{b'} \left\{\Pr(x_b = x_{b'}) - \Pr(x_b \ne x_{b'})\right\} =p_b p_{b'} \phi(b, b').
	\end{align}
	If $x$ is a uniformly distributed, $\phi(b, b')=0$ because $\Pr(x_b)= \frac{1}{2}$ for any $b\in[0, B-1]$. 	
\end{IEEEproof}

Note that \eqref{eq:mse_any} is not convex since the $p_b p_{b'}$ values are not convex and $\phi(b,b')$ can be negative. Fortunately, \eqref{eq:mse_any} can be approximated to \eqref{eq:mse_qfunc} because the second term in the right side of \eqref{eq:mse_any} is much smaller than the first term as shown in the following claim.

\begin{claim}
	If $p_0 \ll \frac{1}{2}$, then \eqref{eq:mse_any} can be approximated as \eqref{eq:mse_qfunc}. 
\end{claim}
\begin{IEEEproof}
	We can rewrite \eqref{eq:mse_any} as follows:
	\begin{align}
	\mathsf{MSE}(x) = p_0 + \sum_{b=1}^{B-1}{(4^b + c_b)p_b} 
	\end{align}
	where $c_b = 2^{b+1} \sum_{b'=0}^{b-1}{2^{b'}p_{b'} \phi(b,b')}$. Hence, 
	\begin{align}
	\left|c_b\right| & \le 2^{b+1} \sum_{b'=0}^{b-1}{2^{b'}p_{b'} \left| \phi(b,b') \right|} \le 2^{b+1} \sum_{b'=0}^{b-1}{2^{b'}p_{b'}} \label{eq:approx_pf1} \\
	& \le 2^{b+1} p_0 \sum_{b'=0}^{b-1}{2^{b'}} \label{eq:approx_pf2} = 2^{b+1}(2^b-1) p_0
	\end{align} 
	where \eqref{eq:approx_pf1} follows from $\left|\phi(b,b')\right|\le 1$. Also, \eqref{eq:approx_pf2} follows from the fact that $p_0 \ge p_b$ for $b \in [1, B-1]$ in our optimization problems.   
	If $4^b \gg 2^{b+1}(2^b-1) p_0$ for every $b\in[1,B-1]$, then we can neglect the MSE difference between a uniformly distributed source and non-uniformly distributed sources, which is satisfied by the condition $p_0 \ll \frac{1}{2}$.  
\end{IEEEproof}

We observe that \eqref{eq:mse_any} is very close to \eqref{eq:mse_qfunc} in many cases even if $p_0 \approx \frac{1}{2}$ (see Table~\ref{tab:psnr} in Section~\ref{sec:numerical}). The reason is that the second term of \eqref{eq:mse_any} cancels out due to sign changes of $\phi(b,b')$. 

\subsection{Non-Gaussian Noise Models}

Although SRAM noise is well-modeled as a Gaussian distribution, the proposed optimization problems can be extended to non-Gaussian noise models. We show that the convexity of proposed optimization problems are maintained if the noise is unimodal and symmetric with zero mean. 

\begin{claim}
	If the noise has a unimodal and symmetric distribution with zero mean, then $\mathsf{MSE}(\vect{\Delta})$ is convex. 
\end{claim}
\begin{IEEEproof}
	Suppose that the noise distribution is $f(t)$, which is a unimodal and symmetric distribution with zero mean. Then, the bit error probability is given by $p_b = \int_{\Delta_b}^{\infty}{f(t) dt}$. Note that $
	\frac{d^2 p_b}{d \Delta_b^2} = -\frac{d f(\Delta_b)}{d \Delta_b} \ge 0$
	which follows from $\frac{d f(\Delta_b)}{d \Delta_b} \le 0$ for $\Delta_b \ge 0$. Since the MSE is the nonnegative weighted sum of bit error probabilities, the MSE is also convex. 
\end{IEEEproof}

Hence, the optimization problems to minimize energy, delay, and EDP for a given constraint on MSE are convex if the noise distribution is unimodal, symmetric, and has zero mean. 

\section{Architecture and Discrete Swings}\label{sec:discrete}

In the previous section, we determined the optimized swings assuming that any real value can be assigned to bit-level swings. However, current SRAM architectures and circuits do not support fine-grained bit-level swing assignments. In this section, we propose an SRAM architecture to enable bit-level swing control. Also, we provide algorithms to optimize discrete-valued swings rather than continuous-valued swings. 

\subsection{Proposed Architecture}

In~\cite{Abu-Rahma2010reducing}, an SRAM architecture that allocates different swings for each memory instance (array or sub-array) was introduced. The fine-grained swings were achieved by WL pulse-width control with little overhead. This architecture attempts to compensate for the impact of spatial variations by applying different pulse-widths to each sub-array. 

\begin{figure}[!t]
	\centering
	\vspace{-3mm}
	\includegraphics[width=0.4\textwidth]{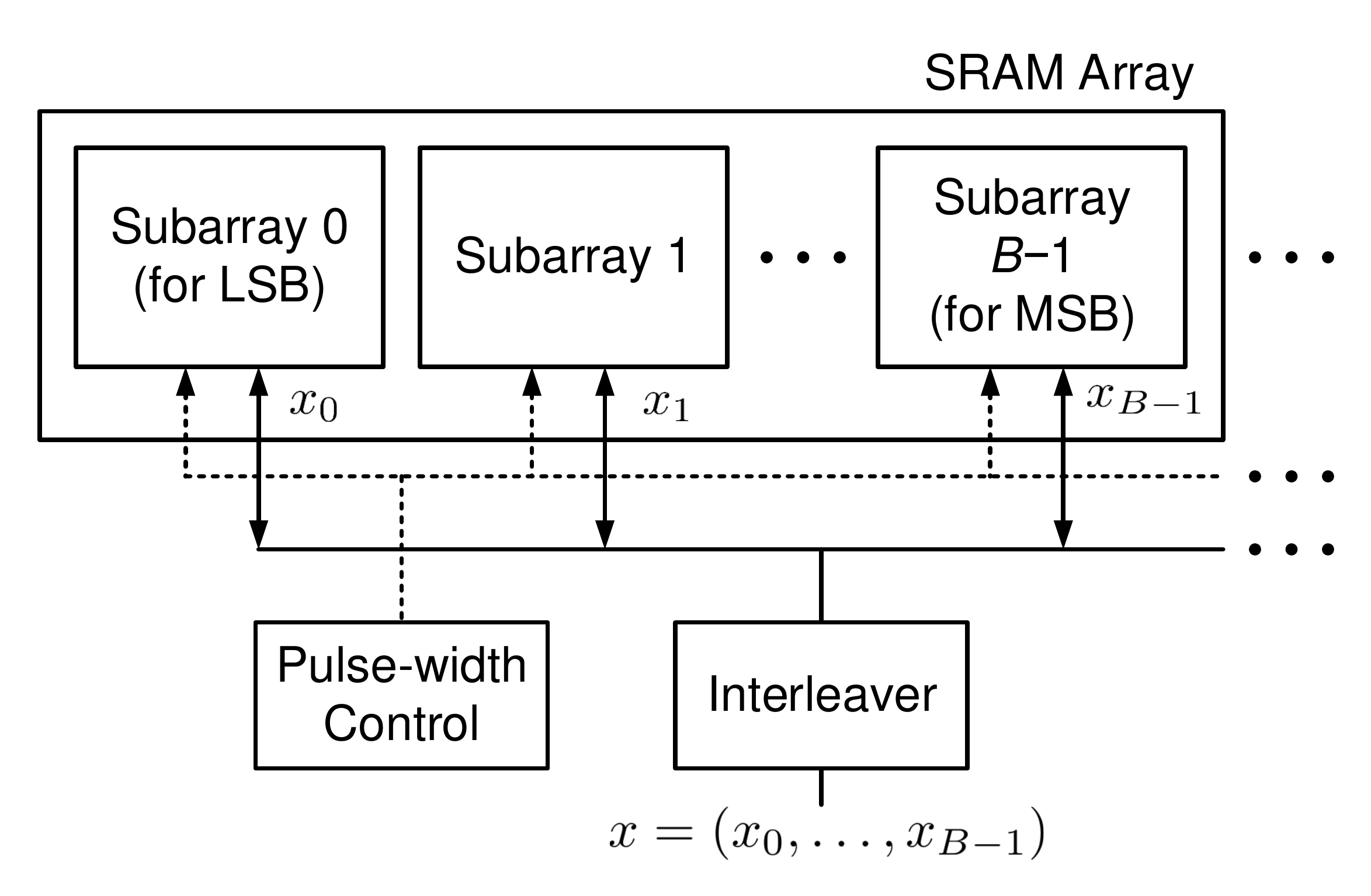}
	\vspace{-8mm}
	\caption{Proposed interleaved architecture.}
	\label{fig:arch}
	\vspace{-8mm}
\end{figure}

By tweaking the architecture of~\cite{Abu-Rahma2010reducing}, we propose an architecture that controls bit-level swings in an efficient manner. We can separate the data for each bit position in different sub-arrays by interleaving (see Fig.~\ref{fig:arch}). Note that interleaving is already used in most SRAMs for soft-error immunity~\cite{Maiz2003char,Osada2003tunnel}. Hence, our architecture does not incur additional overhead, compared to the architecture in \cite{Abu-Rahma2010reducing}.

The proposed architecture enables fine-grained bit-level swing control by adjusting pulse-width for each sub-arrays. Also, dynamic swing control depending on the time-varying fidelity requirement can be achieved by pulse-width control in Fig.~\ref{fig:arch}. Since pulse-width control is usually implemented by cascaded logic gates~\cite{Abu-Rahma2010reducing}, the swing granularity depends on logic gates response time, which is a finite value. Hence, we present optimization algorithms for discrete swings in the following subsection. 

\subsection{Optimization of Discrete Swings: Discrete Water-filling}

\iffalse By taking into account circuit limitations of pulse-width control unit, we propose Greedy algorithms to optimize discrete swings.\fi By leveraging graphical interpretations from Section~\ref{sec:proposed}, we propose optimization algorithms for discrete swings. For Criterion 1 (minimize energy) and Criterion 2 (maximize speed), our algorithm approximates the Levin--Campello algorithm~\cite{Fox1966discrete,Campello1998optimal,Campello1999practical}. The optimization problem of Criterion 3 (minimize EDP) cannot be solved by the Levin--Campello algorithm and so we develop an algorithm based on \emph{sand-pouring} and \emph{water-filling} interpretation and its KKT conditions. 

Suppose that $\beta$ is the granularity in swings. Our discrete water-filling algorithm (Algorithm~\ref{algo:discrete_water}) attempts to obtain the discrete swings minimizing energy or maximizing speed by a greedy approach. The basic idea is to fill the water from the bit position whose temporal water level is the lowest. 

\begin{algorithm}
	\caption{Discrete water-filling for \eqref{eq:min_energy} and \eqref{eq:max_speed0}} \label{algo:discrete_water}
	\begin{algorithmic}[1]
		\State Set ground level $\vect{g} = \left(g_0, \ldots, g_{B-1}\right)$ \label{algo:discrete_water_g} depending on problems
		\State $\vect{\Delta} \gets \vect{0}$ 
		\While{$\mathsf{MSE}(\vect{\Delta}) > \mathcal{V}$} 
		%		\State $w_b \gets g_b + \frac{\Delta_b^2}{2 \sigma^2}$, $\forall b$ \Comment{Update temporal water level}  
		\State $b \gets \underset{b \in [0, B-1]}{\arg\min}~\left\{g_b +\frac{\Delta_b^2}{2 \sigma^2}\right\}$ \label{algo:discrete_water_rule} \Comment{Lowest water level} 
		\State $\Delta_b\gets \Delta_b + \beta$	\Comment{Fill more water}
		\EndWhile
		\State \textbf{return} $\vect{\Delta}$
	\end{algorithmic}
\end{algorithm}

For Criterion 1, the ground level should be $g_b = \log\frac{\sqrt{2\pi}\sigma}{4^b}$ for $b \in [0, B-1]$ as shown in Fig.~\ref{fig:min_energy}\subref{fig:min_energy_a}. For Criterion 2, we set the ground level as $\vect{g} = \vect{0}$, which represents the flat ground level as shown in Fig.~\ref{fig:max_speed}. 

To minimize energy by discrete swings, we tailor the Levin--Campello algorithm by replacing line \ref{algo:discrete_water_rule} in Algorithm~\ref{algo:discrete_water} with
\begin{align}
b & = \underset{b \in [0, B-1]}{\arg\min} \left\{ \mathsf{MSE}(\vect{\Delta} + \beta \vect{e}_b) - \mathsf{MSE}(\vect{\Delta})\right\} % \nonumber \\
%  & = \underset{b \in [0, B-1]}{\arg\min} \left\{4^b \left(Q\left( \frac{\Delta_b + \beta}{\sigma} \right) - Q\left( \frac{\Delta_b}{\sigma} \right)\right) \right\}
\end{align}
where $\mathbf{e}_b$ is a unit vector where $e_b = 1$ and $e_b' = 0$ for $b' \ne b$. Since $\mathsf{MSE}(\vect{\Delta})$ is the sum of convex functions, the discrete swings obtained by the Levin--Campello algorithm are optimal. We show that Algorithm~\ref{algo:discrete_water} is an approximation of the Levin--Campello algorithm.

\begin{corollary}\label{cor:discrete_water}
	The solution by Algorithm~\ref{algo:discrete_water} converges to the solution by Levin--Campello algorithm for small $\beta$. 
\end{corollary}
\begin{IEEEproof}By Lemma~\ref{lemma:mse_qfunc}, 
\begin{align}
\mathsf{MSE}(\vect{\Delta} + \beta \vect{e}_b) - \mathsf{MSE}(\vect{\Delta}) = 4^b \left(Q\left( \frac{\Delta_b + \beta}{\sigma} \right) - Q\left( \frac{\Delta_b}{\sigma} \right)\right). \label{eq:discrete_mse}
\end{align}
As $\beta \rightarrow 0$, \eqref{eq:discrete_mse} converges to 
\begin{equation}\label{eq:discrete_mse_converg}
	\beta \cdot 4^b \cdot \frac{\partial Q\left(\frac{\Delta_b}{\sigma}\right)}{\partial \Delta_b} =- \beta \cdot \frac{4^b}{\sqrt{2\pi}\sigma} \exp{\left(-\frac{\Delta_b^2}{2 \sigma^2}\right)}. 
\end{equation}
We can consider choosing $b$ that minimizes \eqref{eq:discrete_mse_converg} as follows:
\begin{align}
b & = \arg\min \left\{ - \beta \cdot \frac{4^b}{\sqrt{2\pi}\sigma} \exp{\left(-\frac{\Delta_b^2}{2 \sigma^2}\right)}\right\} = \arg\min \left\{ \log{\frac{\sqrt{2\pi}\sigma}{4^b}} + \frac{\Delta_b^2}{2 \sigma^2}  \right\}, 
\end{align}
which is equivalent to line~\ref{algo:discrete_water_rule} of Algorithm~\ref{algo:discrete_water}. 
\end{IEEEproof}
Numerical results in Section~\ref{sec:numerical} show that the discrete swings obtained by Algorithm~\ref{algo:discrete_water} are almost identical to the solutions by the Levin--Campello algorithm.  

We present an algorithm to obtain discrete swings to minimize EDP in Algorithm~\ref{algo:discrete_sand_water}. The Levin-Campello algorithm cannot solve this problem since the $\rho = \max{(\vect{\Delta})}$ in EDP cannot be handled by the Levin-Campello algorithm. By leveraging the sand-pouring and water-filling interpretation of Fig.~\ref{fig:min_edp} and KKT conditions, Algorithm~\ref{algo:discrete_sand_water} attempts to pour sand and fill water iteratively. 

\begin{algorithm}
	\caption{Sand-pouring and discrete water-filling for \eqref{eq:min_edp}} \label{algo:discrete_sand_water}
	\begin{algorithmic}[1]
		\State $g_b \gets \log\frac{\sqrt{2\pi}\sigma}{4^b}$ for all $b \in [0, B-1]$ \Comment{Set ground level}				
		\State $\vect{\Delta} \gets \vect{0}$, $\vect{\eta} \gets \vect{0}$, $\vect{s} \gets \vect{0}$
		\While{$\mathsf{MSE}(\vect{\Delta}) > \mathcal{V}$} 
		\State $\rho \gets \max(\vect{\Delta})$		
		\State $b \gets \underset{b \in [0, B-1]}{\arg\min}~\left\{g_b + s_b\right\}$ \label{algo:sand_position} \Comment{Lowest sand level}
		\State $\eta_b \gets \eta_b + \beta$ \label{algo:sand_increase} \Comment{Pour more sand}
		\For{$b=0$ to $B-1$}
		\State $s_b \gets \log\left(1 + \frac{\eta_b}{\rho}\right)$ \Comment{Calculate sand depth}
		\EndFor
		\State $b \gets \underset{b \in [0, B-1]}{\arg\min}~\left\{g_b + s_b + \frac{\Delta_b^2}{2 \sigma^2}\right\}$ \label{algo:water_position}   \Comment{Lowest water level}
		\State $\Delta_b\gets \Delta_b + \beta$ \label{algo:water_increase}	\Comment{Fill more water}
		\EndWhile
		\State \textbf{return} $\vect{\Delta}$
	\end{algorithmic}
\end{algorithm}

At each iteration, Algorithm~\ref{algo:discrete_sand_water} first pours more sand from the lowest sand level as shown in line~\ref{algo:sand_position}. Note that the sand level of each bit position is the sum of the corresponding ground level and sand depth. We increase $\eta_b$ by $\beta$ in line~\ref{algo:sand_increase} and $\Delta_b$ by $\beta$ in line~\ref{algo:water_increase} at each iteration to satisfy the optimal condition $\sum{\eta_b} = \sum{\Delta_b}$ (see~\eqref{eq:cr3_KKT_gr_2} in Appendix~\ref{pf:min_edp}). After increasing $\eta_b$, the sand depth $s_b$ of each bit position is calculated by Corollary~\ref{cor:sand_depth}, which indicates the increased amount of sand. Afterwards, water is filled from the bit position whose water level is the lowest. Note that the sand depth $s_b$ affects the water level  unlike Algorithm~\ref{algo:discrete_water} (Compare line \ref{algo:discrete_water_rule} of Algorithm~\ref{algo:discrete_water} and line \ref{algo:water_position} of Algorithm~\ref{algo:discrete_sand_water}). 

Numerical results in Section~\ref{sec:numerical} show that the EDP loss due to discrete swings of Algorithm~\ref{algo:discrete_sand_water} is negligible for moderate granularity $\beta$. 

\section{Numerical Results}\label{sec:numerical}

\begin{figure}[t]
	\vspace{-3mm}
	\centering
	\subfloat[]{\includegraphics[width=0.4\textwidth]{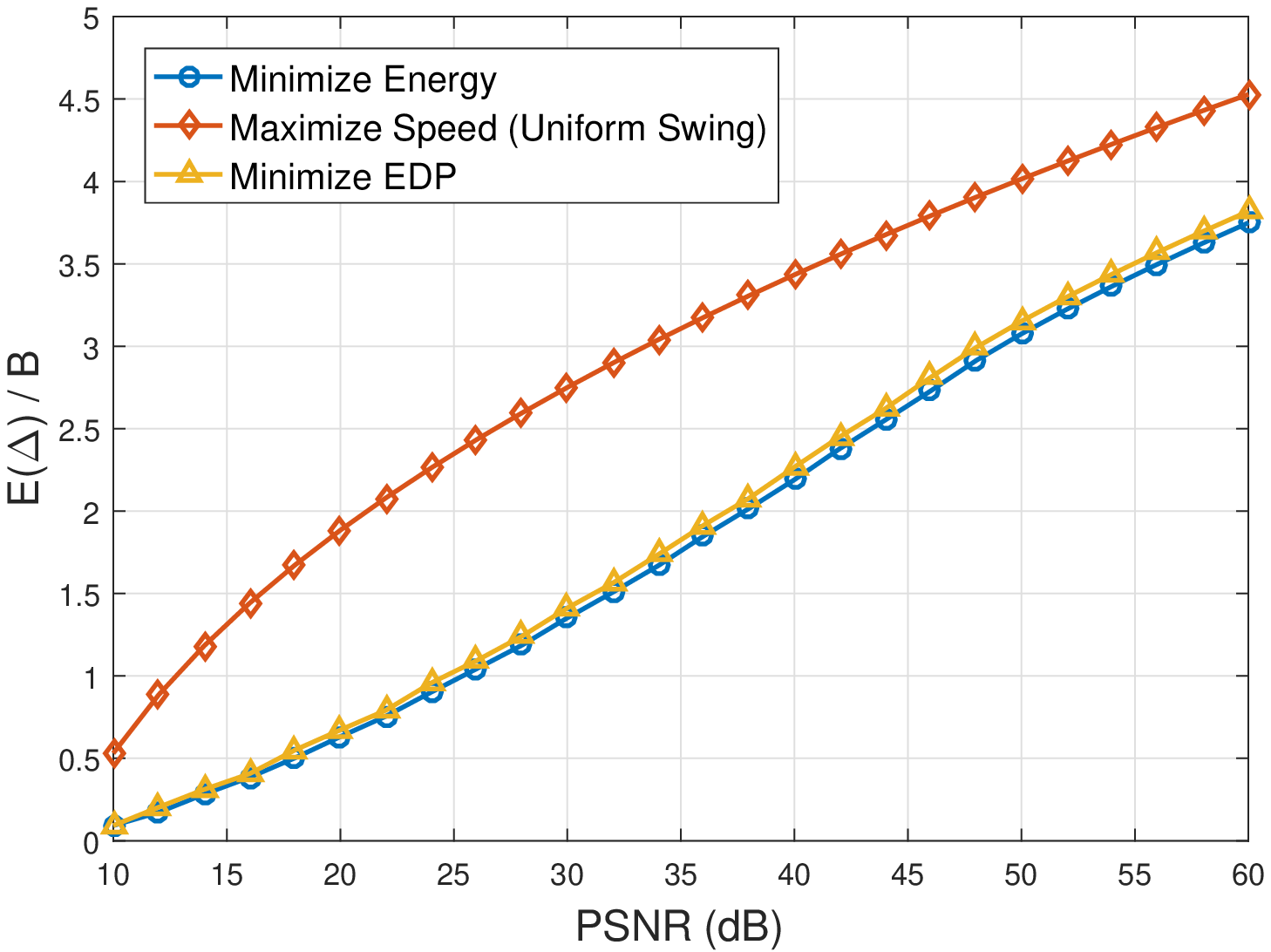}
		\label{fig:plot_energy_a}}
	\hfill
	\vspace{-1mm}
	\subfloat[]{\includegraphics[width=0.4\textwidth]{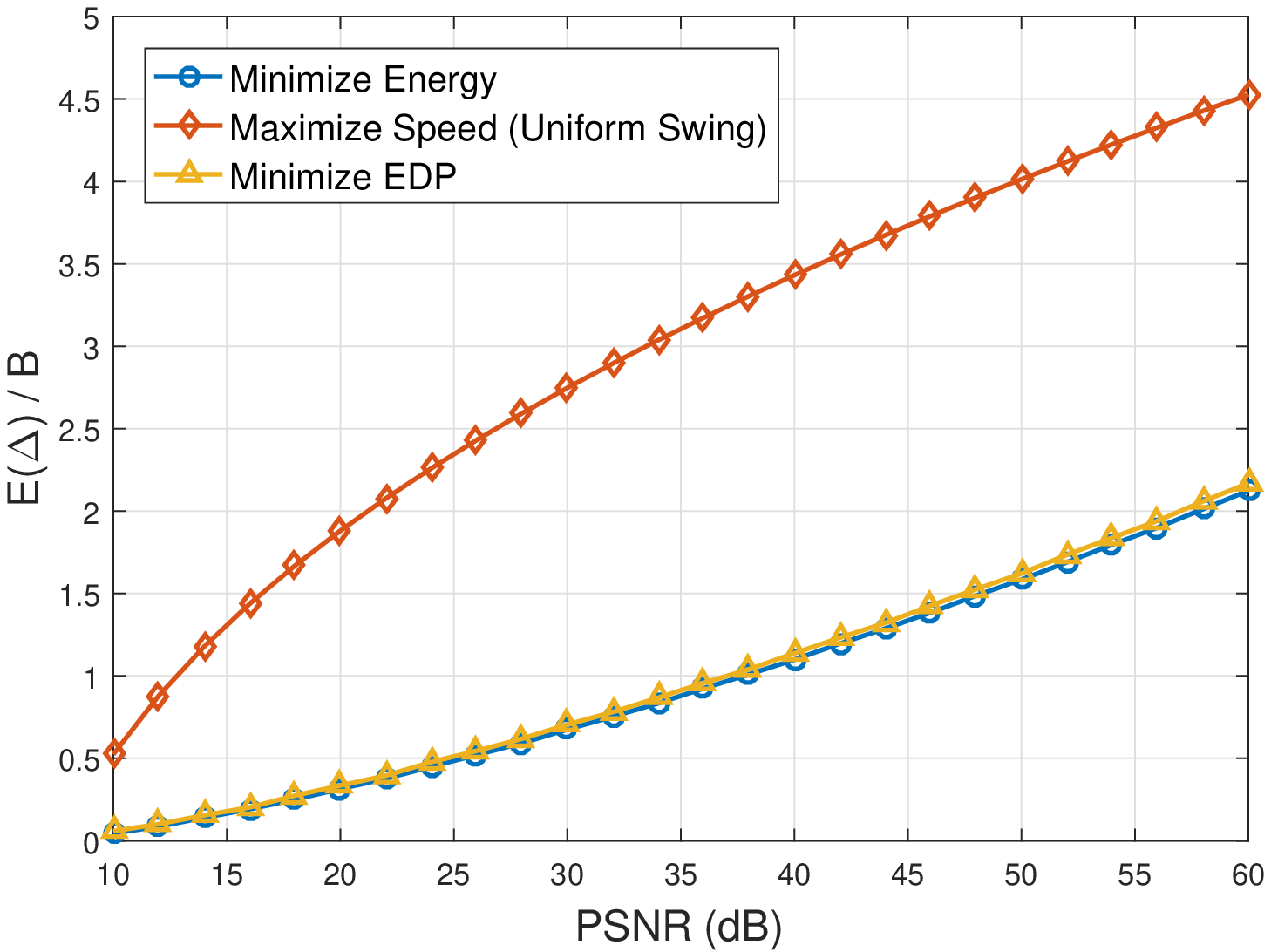}
		\label{fig:plot_energy_b}}
	\caption{Comparison of energy consumption for (a) $B=8$ and (b) $B=16$ ($\sigma$ = 1).}
	\label{fig:plot_energy}
	\vspace{-8mm}
\end{figure}

\begin{figure}[t]
	\centering
	\subfloat[]{\includegraphics[width=0.4\textwidth]{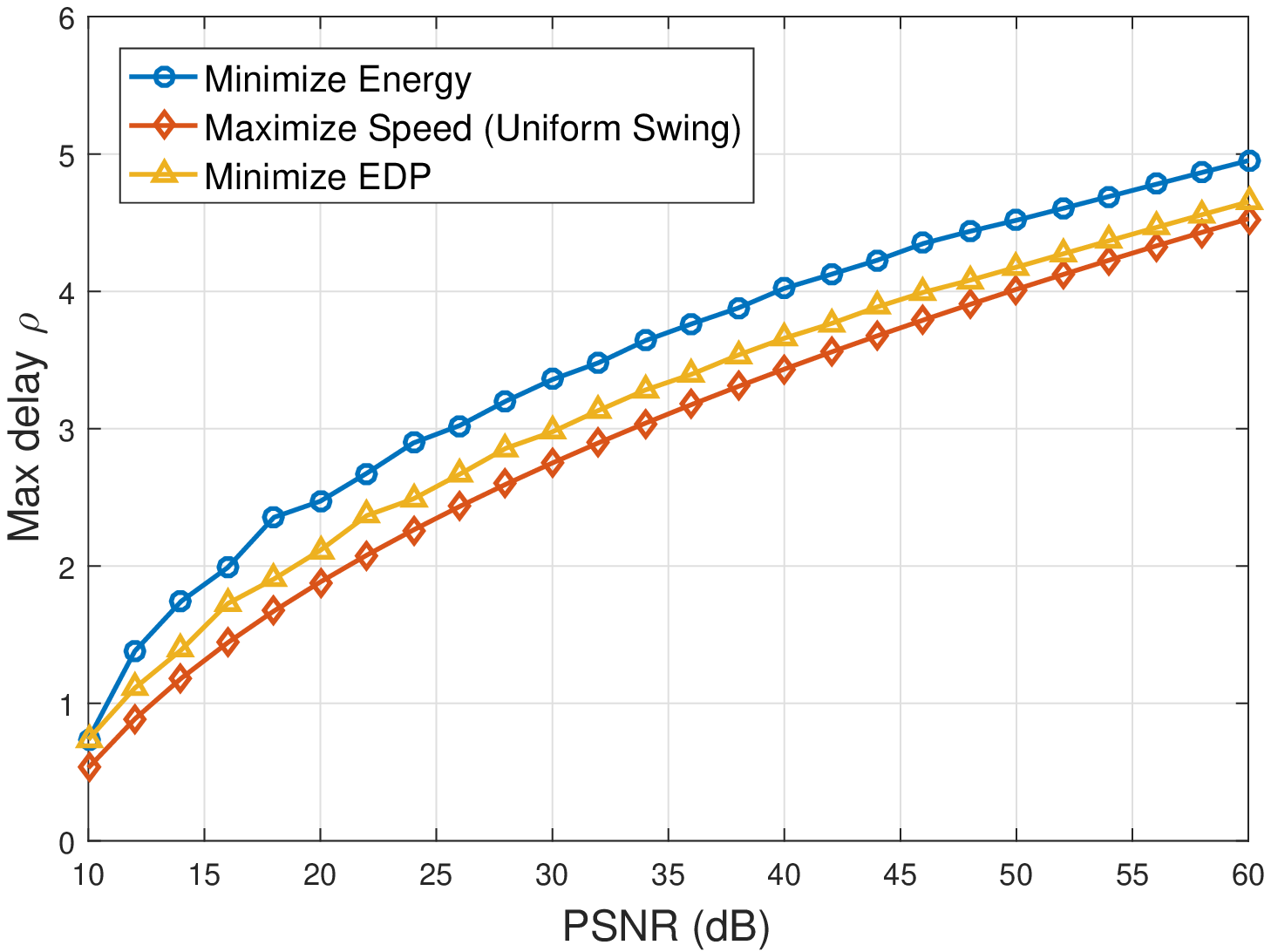}
		\label{fig:plot_delay_a}}
	\hfill
	%	\vspace{-3mm}
	\subfloat[]{\includegraphics[width=0.4\textwidth]{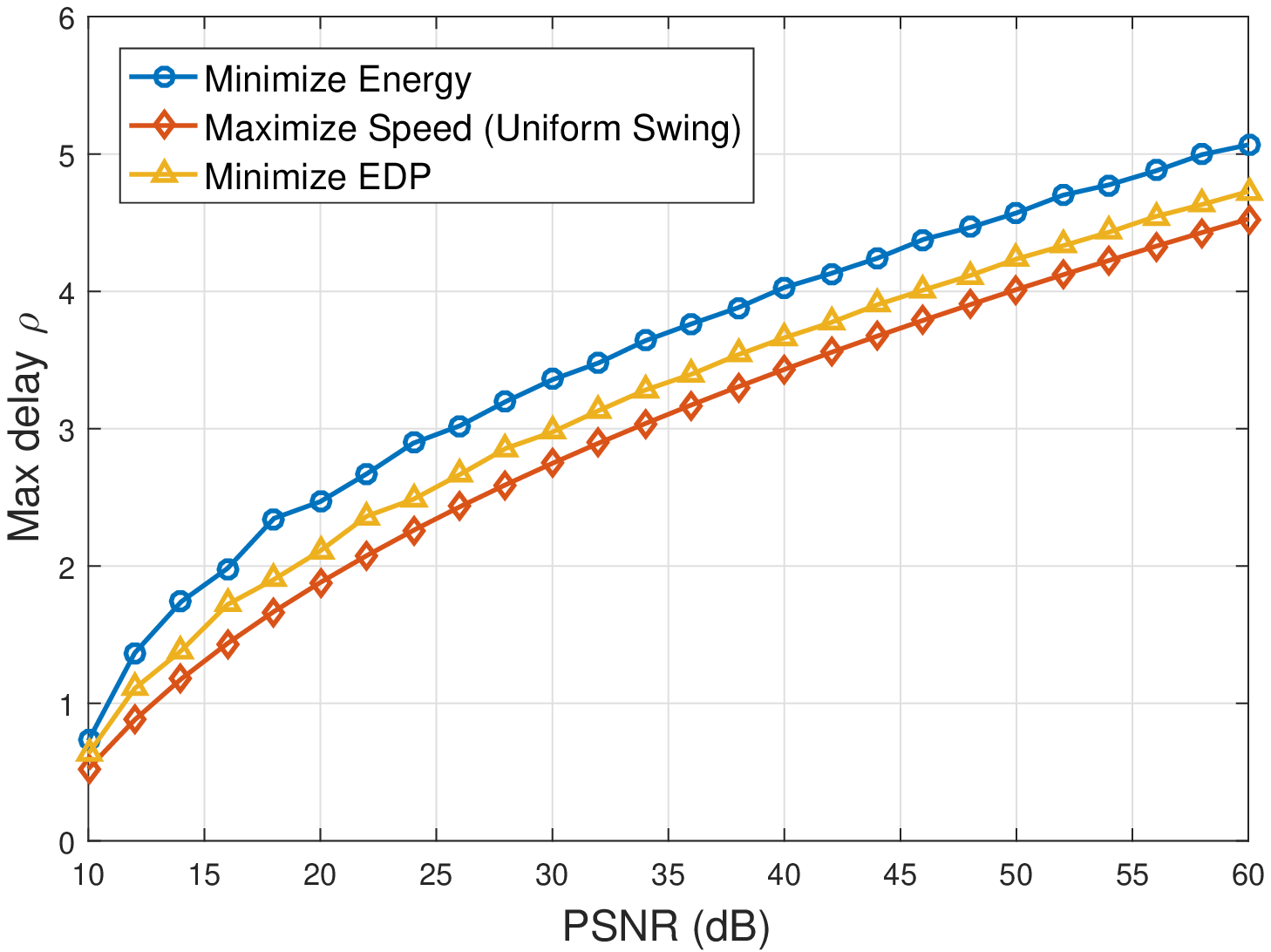}
		\label{fig:plot_delay_b}}
	\caption{Comparison of maximum delay for (a) $B=8$ and (b) $B=16$  ($\sigma$ = 1).}
	\label{fig:plot_delay}
	\vspace{-8mm}
\end{figure}

\begin{figure}[t]
	\centering
	\subfloat[]{\includegraphics[width=0.4\textwidth]{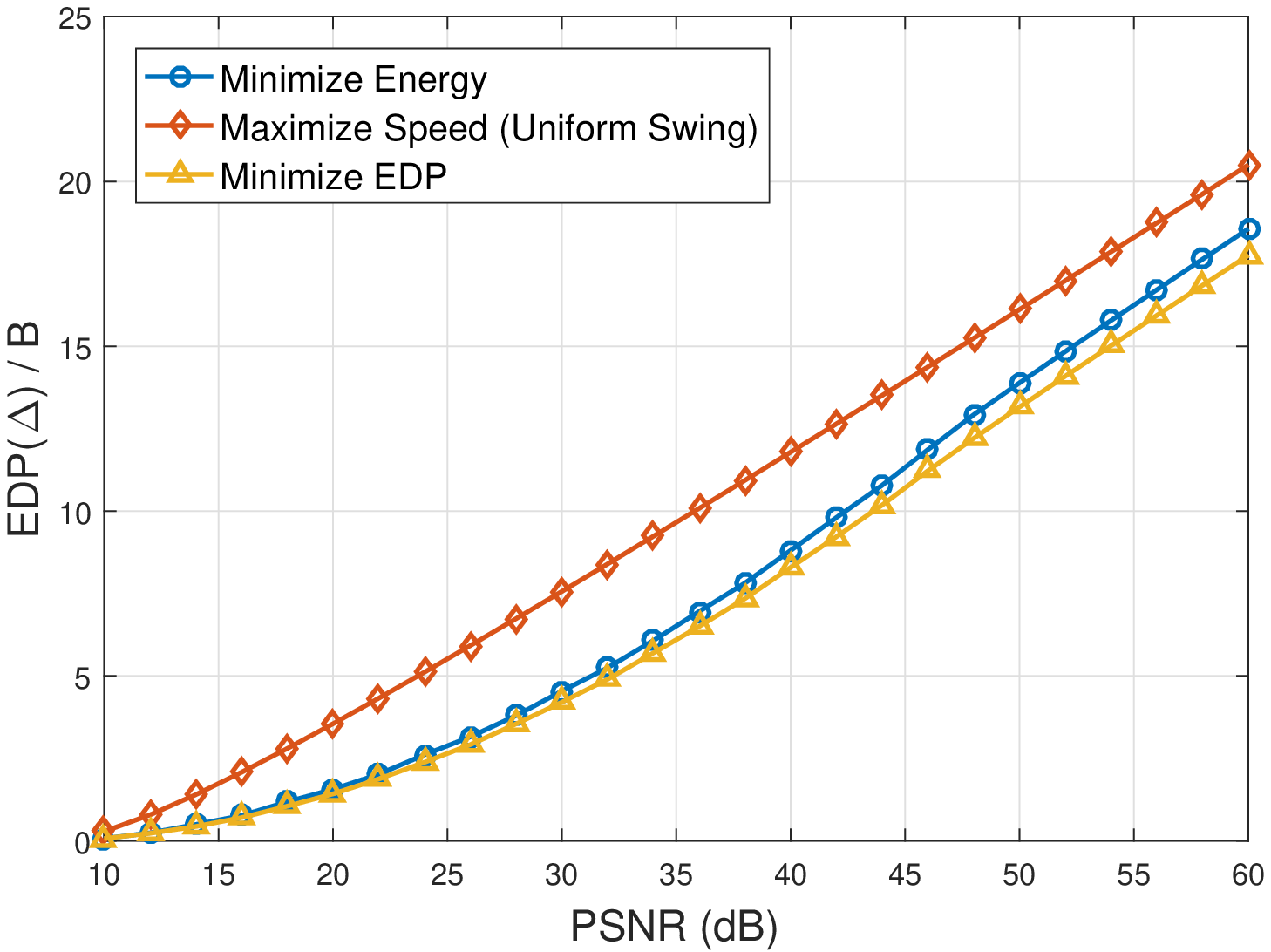}
		\label{fig:plot_edp_a}}
	\hfill
	%	\vspace{-3mm}
	\subfloat[]{\includegraphics[width=0.4\textwidth]{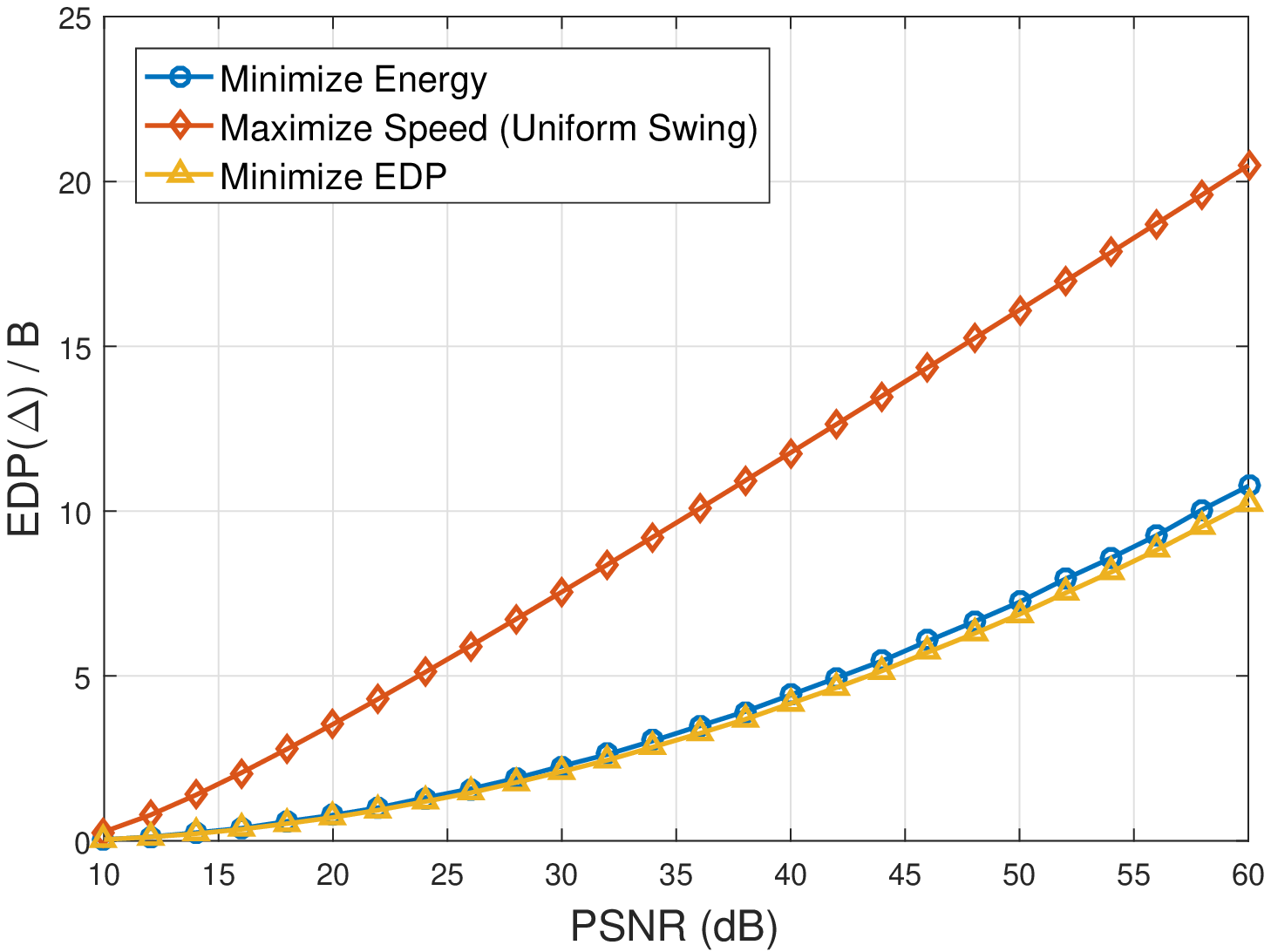}
		\label{fig:plot_edp_b}}
	\caption{Comparison of EDP for (a) $B=8$ and (b) $B=16$ ($\sigma$ = 1).}
	\label{fig:plot_edp}
	\vspace{-8mm}
\end{figure}

\begin{figure}[t]
	\centering
	\subfloat[]{\includegraphics[width=0.4\textwidth]{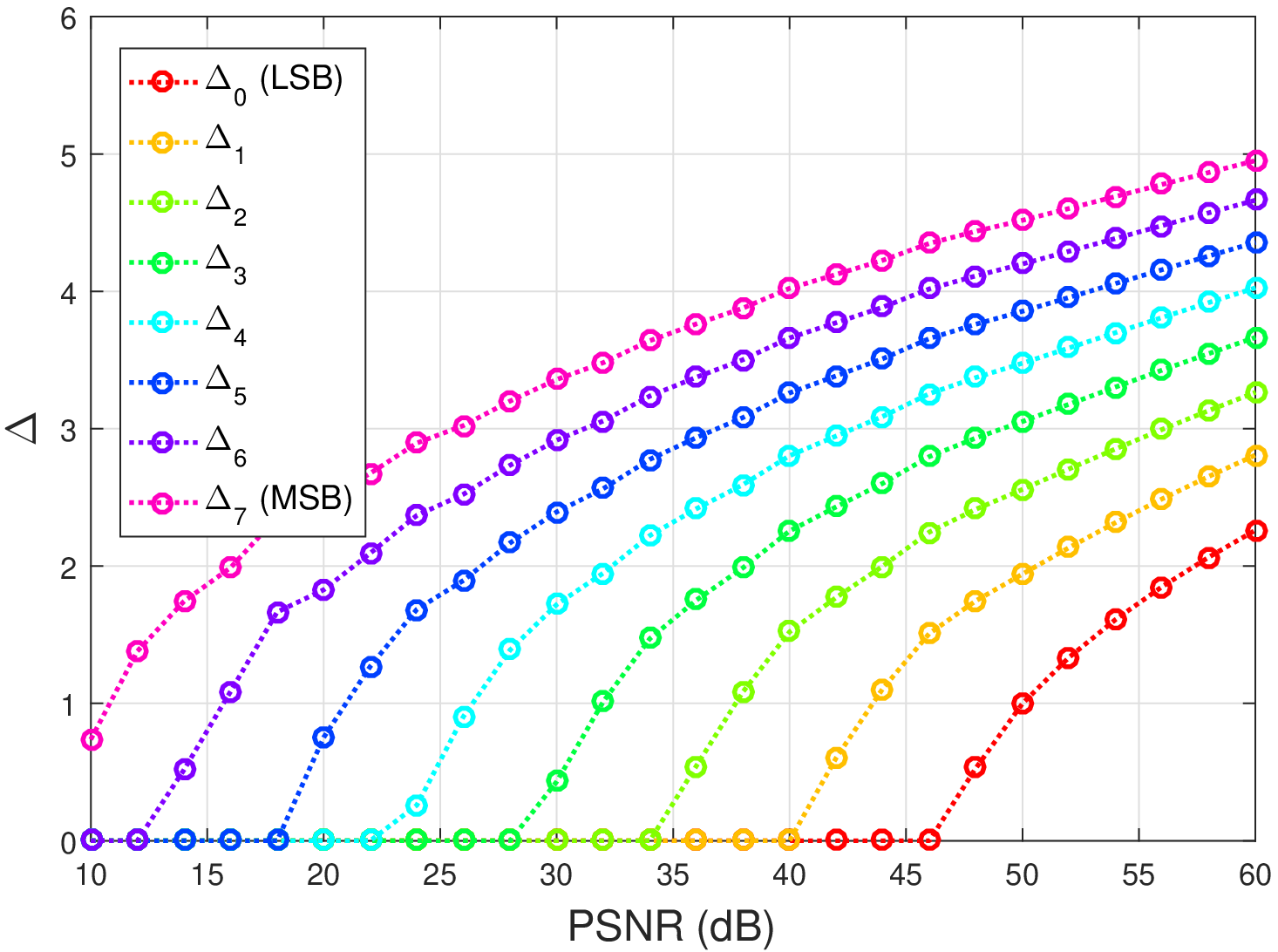}
		\label{fig:plot_sol_B8_1}}
	\hfil
	%	\vspace{-3mm}
	\subfloat[]{\includegraphics[width=0.4\textwidth]{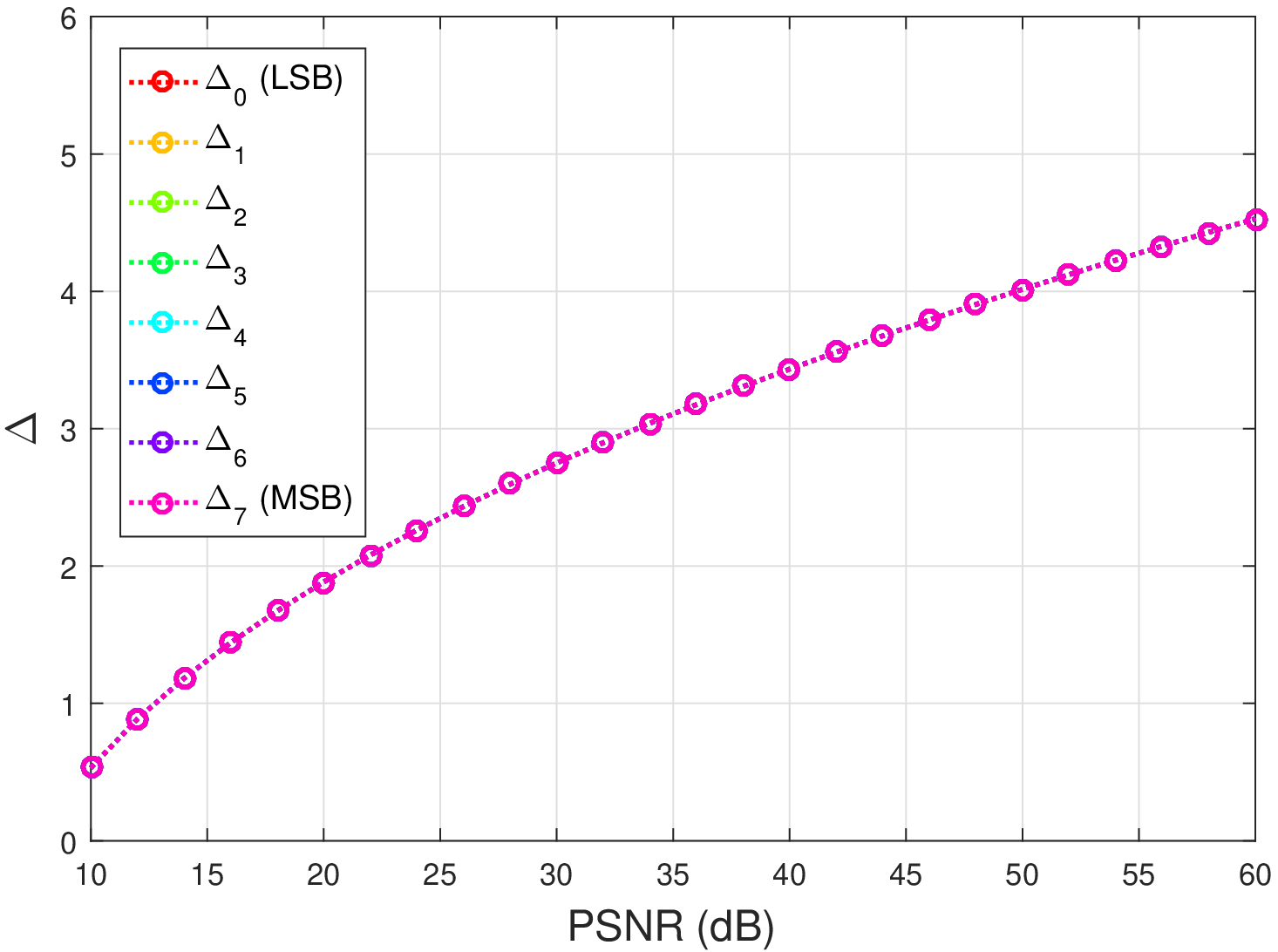}
		\label{fig:plot_sol_B8_2}}
	\hfil
	%   \vspace{-3mm}
	\subfloat[]{\includegraphics[width=0.4\textwidth]{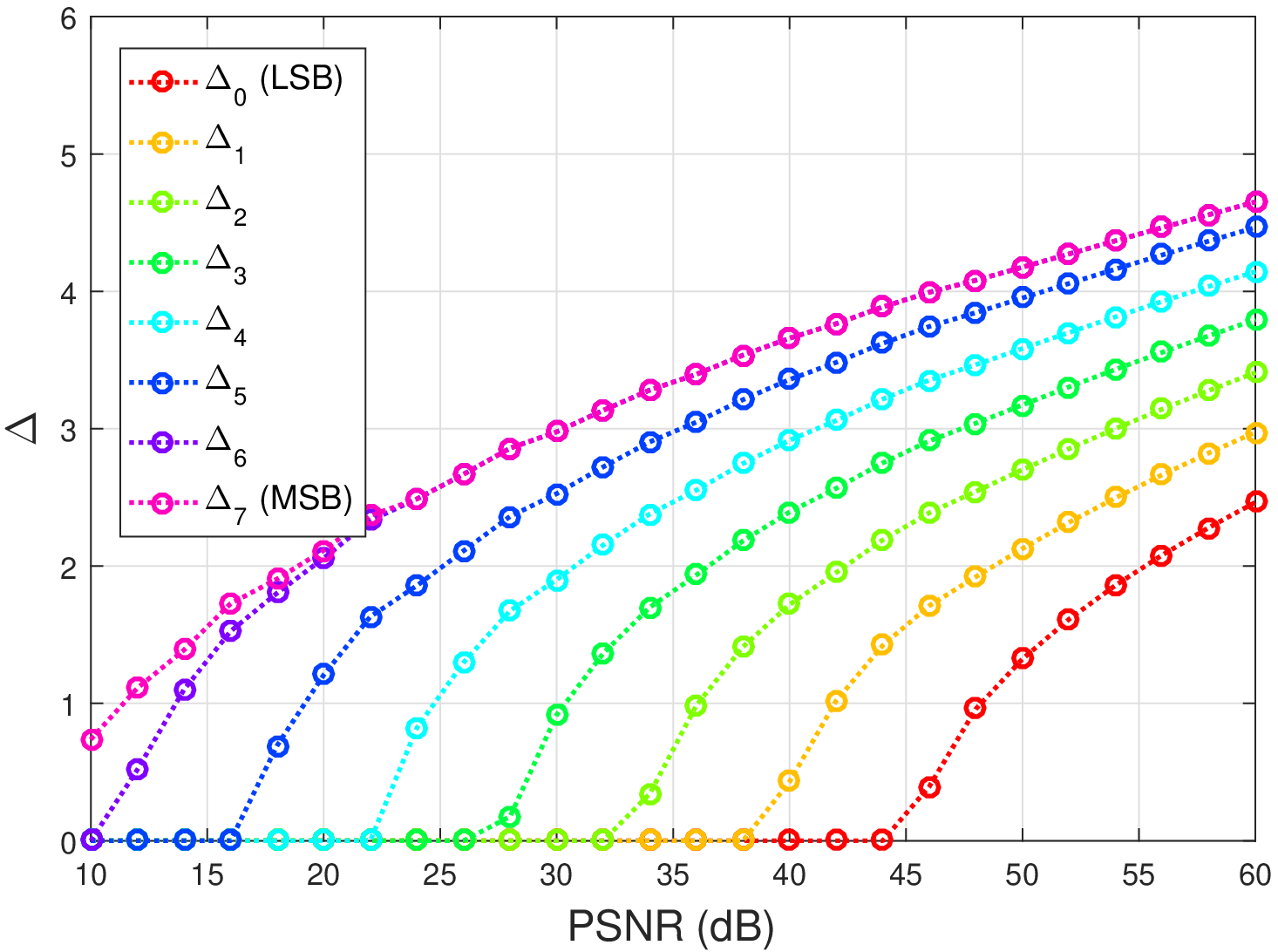}
		\label{fig:plot_sol_B8_3}}	
	\caption{Optimal solutions (a) minimizing energy, (b) maximizing speed, and (c) minimizing EDP ($\sigma$ = 1).}	
	\label{fig:plot_sol_B8}
	\vspace{-8mm}
\end{figure}

We evaluate the solutions of the three optimization problems for both continuous and discrete swings. Note that the solution of maximizing speed is equivalent to the conventional uniform swing as noted in Remark~\ref{remark:max_speed}. Also, we compare the proposed optimization solution to LSB dropping and selective ECCs. 

Fig.~\ref{fig:plot_energy} compares the read energy consumption $\mathsf{E}(\vect{\Delta})$ as in Definition~\ref{def:energy} for a given constraint of peak signal-to-noise ratio (PSNR). The PSNR depends on the MSE as
\begin{equation}
\mathsf{PSNR} = 10\log_{10}{\frac{\left(2^B-1\right)^2}{\mathsf{MSE}(\vect{\Delta})}}.
\end{equation}
At PSNR = 30dB, the optimal solution of~\eqref{eq:min_energy} (i.e., minimizing energy) reduces the energy consumption by half for $B=8$, compared to uniform swing (i.e., maximizing speed). For $B=16$, the energy consumption of energy-optimal swing will be only quarter, compared to the uniform swing. Note that energy consumption of EDP-optimal swing is slightly worse than that of energy-optimal swing. 

Fig.~\ref{fig:plot_delay} compares the maximum delay $\rho$ as in Definition~\ref{def:max_swing} for a given PSNR. The conventional uniform swing minimizes the maximum delay; hence it is the speed-optimal solution. The swings minimizing energy achieve significant energy savings at the cost of speed (e.g., the maximum delay increase of 20\% at PSNR = 30dB). The EDP-optimal swings increase only 8\% of maximum delay at PSNR = 30dB. 

Fig.~\ref{fig:plot_edp} compares the EDP for a given PSNR. As formulated, the swings minimizing EDP show the best results. The EDP can be reduced by 45\% for $B=8$ at PSNR = 30dB. The EDP improvement will be much more for $B=16$, e.g., 75\% EDP saving at PSNR = 30dB. Note that slight loss of speed performance can result in significant energy and EDP savings. 

Fig.~\ref{fig:plot_sol_B8} shows optimal solutions to (a) minimize energy, (b) minimize maximum delay, and (c) minimize EDP. As shown in Fig.~\ref{fig:plot_sol_B8}\subref{fig:plot_sol_B8_1}, we should allocate larger swings for more significant bits. Also, we observe that the swings for several LSBs can be zero depending on PSNR, e.g., $\Delta_0 = \Delta_1 = \Delta_2 = 0$ at PSNR = 30dB, a refined kind of LSB dropping. These numerical solutions confirm Theorem~\ref{thm:min_energy} and its water-filling interpretation in Fig.~\ref{fig:min_energy}. Fig.~\ref{fig:plot_sol_B8}\subref{fig:plot_sol_B8_2} shows the solutions minimizing maximum delay. As we showed in Theorem~\ref{thm:max_speed}, uniform swings minimize the maximum delay. The optimized swings in Fig.~\ref{fig:plot_sol_B8}\subref{fig:plot_sol_B8_3} minimize the EDP. Although the EDP-optimal swings are similar to the energy-optimal swings, we observe that $\Delta_6 = \Delta_7 = \rho$ at PSNR = 30dB. It is because these two bit positions are filled with sand to suppress the maximum delay as shown in Theorem~\ref{thm:min_edp} and its graphical interpretation in Fig.~\ref{fig:min_edp}. 

\begin{figure}[!t]
	\centering
	\subfloat[]{\includegraphics[width=0.4\textwidth]{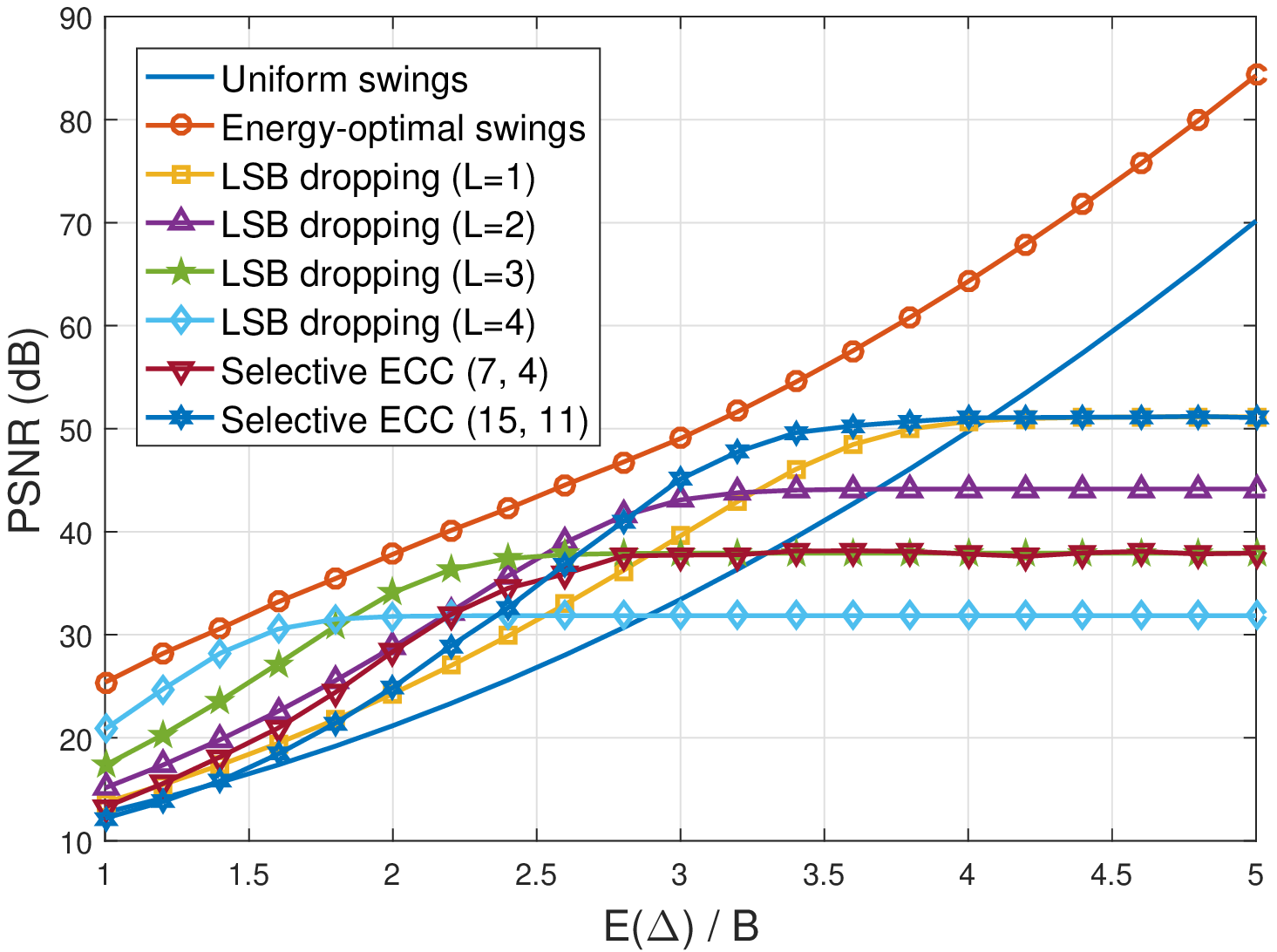}
		\label{fig:plot_LSB_SECC_B8}}
	\hfill
	\subfloat[]{\includegraphics[width=0.4\textwidth]{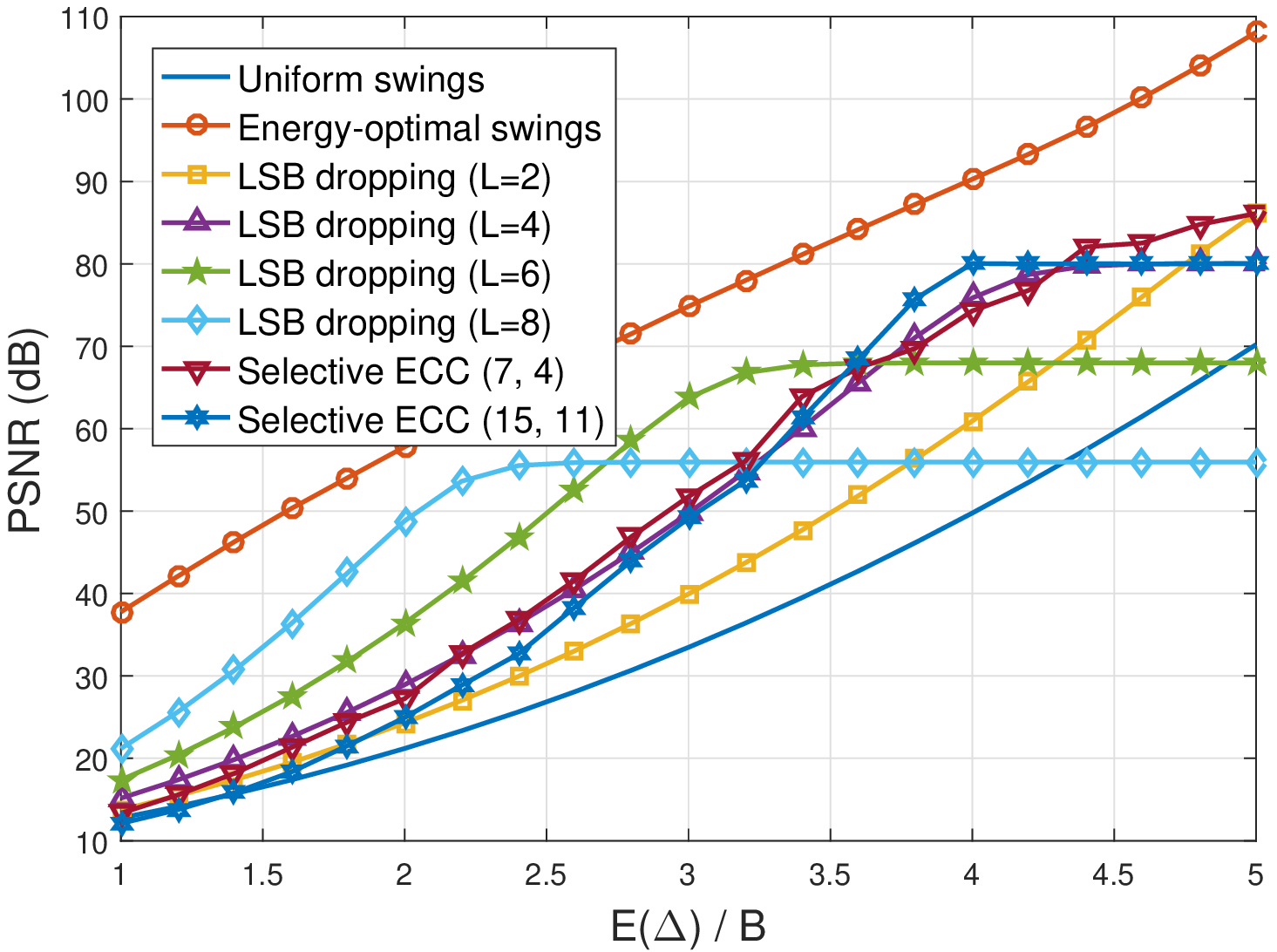}
		\label{fig:plot_LSB_SECC_B16}}
	\vspace{-3mm}
	\caption{Comparison of uniform swings, energy-optimal swings, LSB dropping, and selective ECC (a) $B=8$ and (b) $B=16$ ($\sigma$ = 1).}
	\label{fig:plot_LSB_SECC}
	\vspace{-8mm}
\end{figure}

Fig.~\ref{fig:plot_LSB_SECC} compares uniform swings, energy-optimal swings (i.e., the optimal solutions of \eqref{eq:min_energy}), LSB dropping, and selective ECCs. The proposed energy-optimal swings outperform the other techniques since the energy-optimal swings achieve the target PSNR with the minimum energy $\mathsf{E}(\vect{\Delta})$. 

LSB dropping deactivates $L$ LSBs and allocates uniform swings for $(B-L)$ undropped bit positions. In the low PSNR regime, dropping more LSBs (i.e., larger $L$) can be effective. However, larger $L$ will limit the levels of achievable PSNRs. 

Selective ECCs store parity bits in LSBs to prevent the additional memory overhead. Unlike LSB dropping, selective ECCs allocate uniform swings for all the bit positions. In spite of the LSB information loss, the overall PSNR can be improved by correcting errors in MSBs. As in~\cite{Frustaci2016approximate}, we consider $(n, k)$ Hamming codes for selective ECCs since complicated ECCs are impractical for SRAMs. In a selective ECC (7, 4) for $B=8$, the bits of ($x_7, x_6, x_5, x_4$) are protected by losing information of $(x_2, x_1, x_0)$. Since three LSBs are lost, the PSNR of selective ECC (7, 4) converges to the PSNR by LSB dropping $(L=4)$ as shown in Fig.~\ref{fig:plot_LSB_SECC}\subref{fig:plot_LSB_SECC_B8}. For $B=8$, a (15, 11) Hamming code cannot be incorporated into an 8-bit word. Hence, we store four parity bits of a Hamming (15, 11) codeword in the last LSBs of four different 8-bit words as proposed in~\cite{Frustaci2016approximate}. Note that selective ECC (15, 11) for $B=8$ converges to LSB dropping $(L=1)$ for high $\mathsf{E}(\vect{\Delta})$ since both schemes discard only the last LSBs. In Fig.~\ref{fig:plot_LSB_SECC}\subref{fig:plot_LSB_SECC_B16}, all selective ECCs are applied to one 16-bit word. 

Table~\ref{tab:psnr} compares the PSNRs for uniformly distributed source to real image data (non-uniformly distributed sources) from~\cite{usc-sipi}. Although $p_0 = \frac{1}{2}$ at PSNR = 20dB (see Fig.~\ref{fig:plot_sol_B8}\subref{fig:plot_sol_B8_1} and~\subref{fig:plot_sol_B8_3}), we can observe that their PSNRs are almost the same as the PSNRs of uniformly distributed sources as discussed in Section~\ref{subsec:nonuniform}.

\begin{table*}[t]
	\renewcommand{\arraystretch}{1.1}
	\caption{Comparison of PSNRs [dB] of Uniformly Distributed Sources and Real Image Data}
	\vspace{-3mm}
	\label{tab:psnr}
	\centering	{
		\hfill{}
		\scalebox{0.8}{
		\begin{tabular}{|c|c|c|c|c|c|c|c|c|c|}	\hline
			PSNR of & \multicolumn{3}{c|}{PSNR of Airport} & \multicolumn{3}{c|}{PSNR of Fishing Boat} & \multicolumn{3}{c|}{PSNR of Man} \\ \cline{2-10}
			uniform source & Min energy & Max speed & Min EDP& Min energy & Max speed & Min EDP & Min energy& Max speed & Min EDP \\ \hline \hline
			20 & 19.99& 20.05& 20.07& 20.19& 20.07& 20.18& 19.75& 20.01& 19.85\\ \hline
			24 & 24.05& 24.01& 24.05& 24.10& 24.06& 24.08& 23.82& 24.00& 23.91\\ \hline
			28 & 28.04& 28.01& 28.03& 28.06& 28.02& 28.09& 27.90& 28.00& 27.95\\ \hline
			32 & 32.00& 32.03& 32.03& 32.04& 31.96& 32.02& 31.95& 32.02& 31.98\\ \hline
			36 & 36.00& 36.00& 35.97& 36.03& 36.05& 36.03& 35.99& 35.97& 36.00\\ \hline
			40 & 40.01& 39.95& 39.95& 40.02& 40.05& 40.07& 40.01& 40.03& 40.03\\ \hline
		\end{tabular}}}
	\hfill{}
	\vspace{-8mm}
\end{table*} 

\begin{figure}[t]
	\centering
	\subfloat[]{\includegraphics[width=0.4\textwidth]{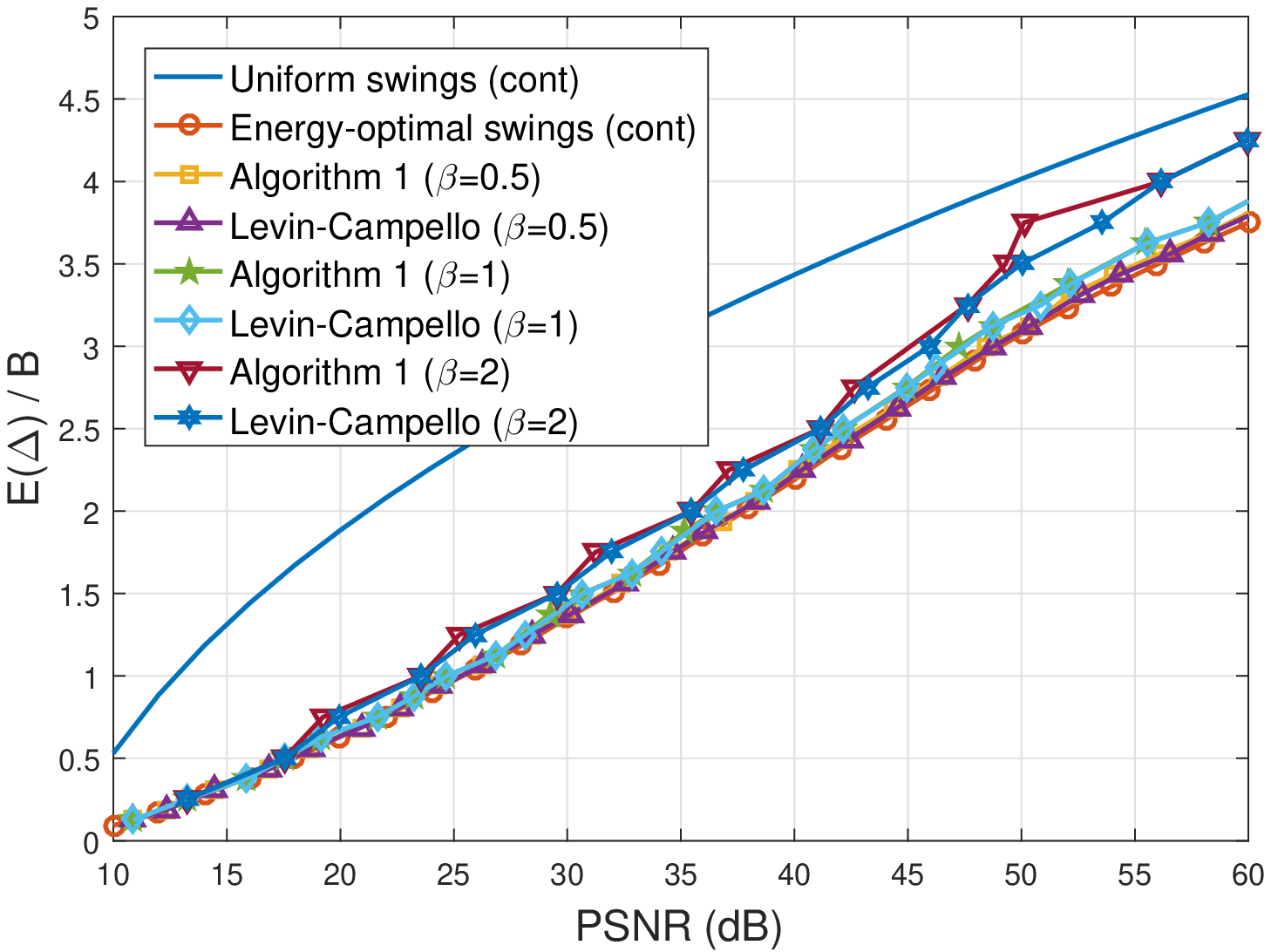}
		\label{fig:plot_energy_discrete_a}}
	\hfill
	%	\vspace{-3mm}
	\subfloat[]{\includegraphics[width=0.4\textwidth]{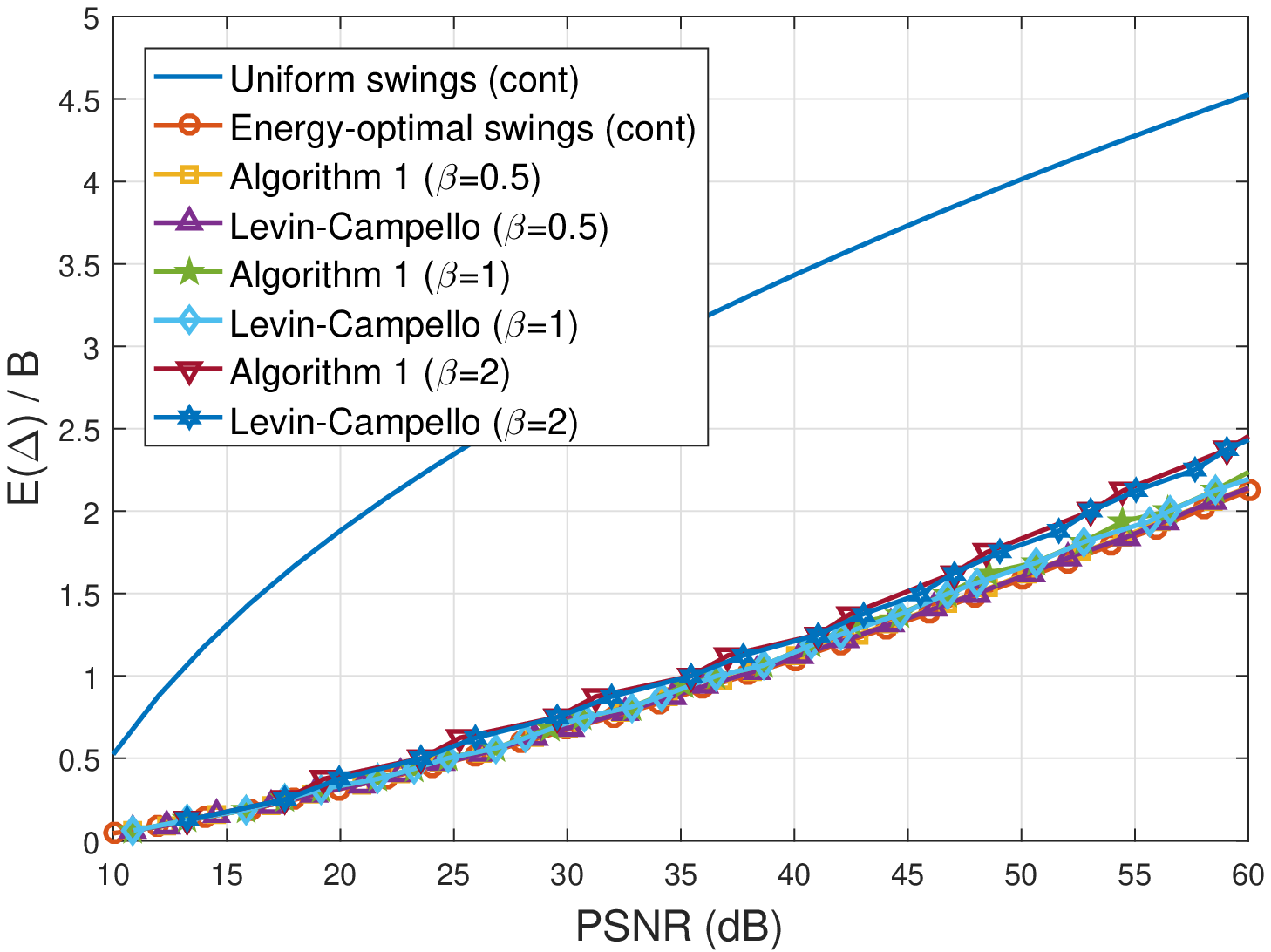}
		\label{fig:plot_energy_discrete_b}}
	\vspace{-3mm}
	\caption{Energy consumption of discrete swings obtained by Algorithm~\ref{algo:discrete_water} and the Levin--Campello algorithm for (a) $B=8$ and (b) $B=16$ ($\sigma$ = 1).}
	\label{fig:plot_energy_discrete}
	\vspace{-8mm}
\end{figure}

Fig.~\ref{fig:plot_energy_discrete} shows that the energy penalty due to discrete swings is negligible for moderate granularity $\beta$. Energy consumption of discrete swings obtained by our Algorithm~\ref{algo:discrete_water} is almost the same as the Levin--Campello algorithm as explained in Corollary~\ref{cor:discrete_water}. 

\begin{figure}[!t]
	\centering
	\subfloat[]{\includegraphics[width=0.4\textwidth]{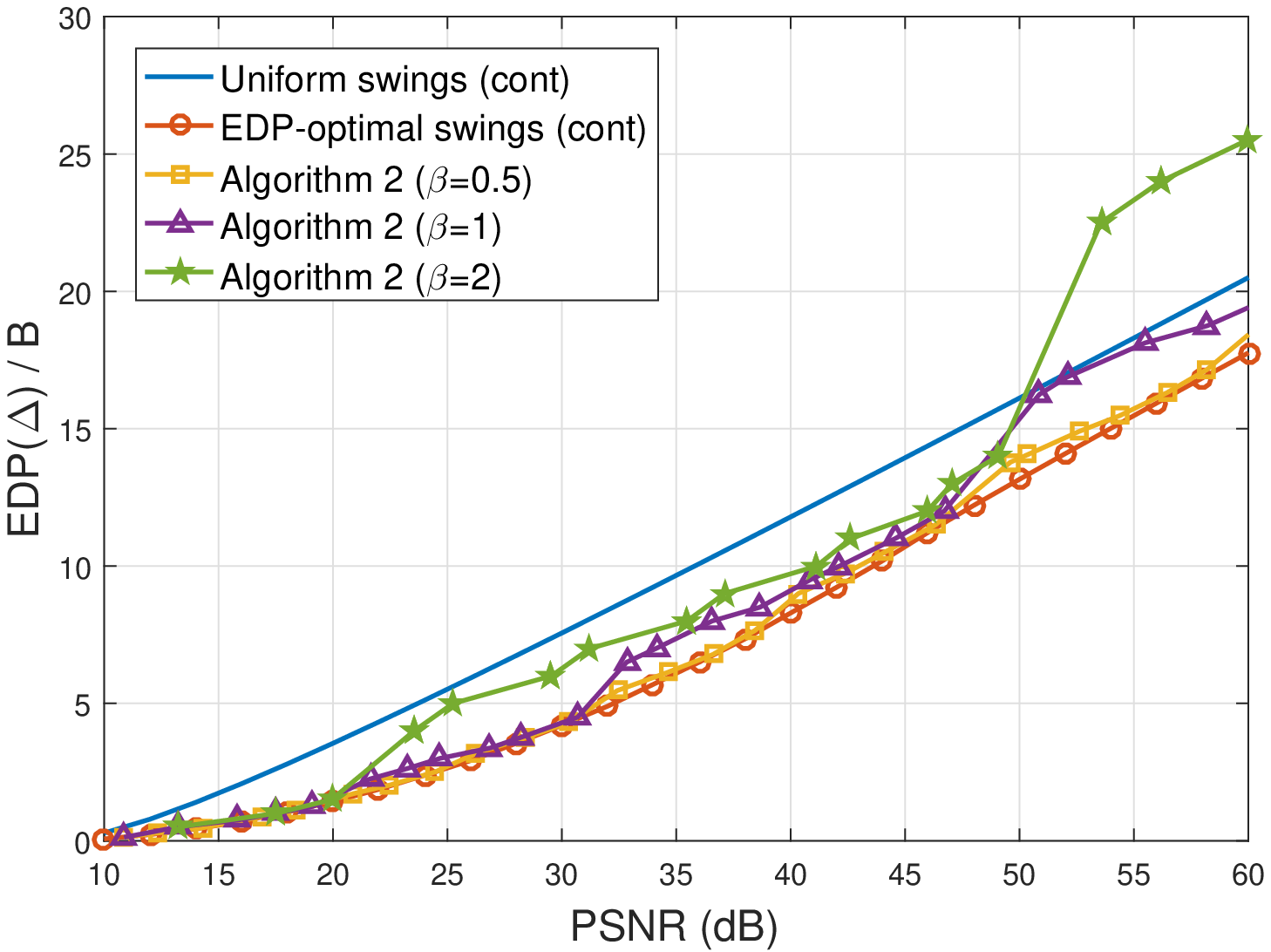}
		\label{fig:plot_edp_discrete_a}}
	\hfill
	%	\vspace{-3mm}
	\subfloat[]{\includegraphics[width=0.4\textwidth]{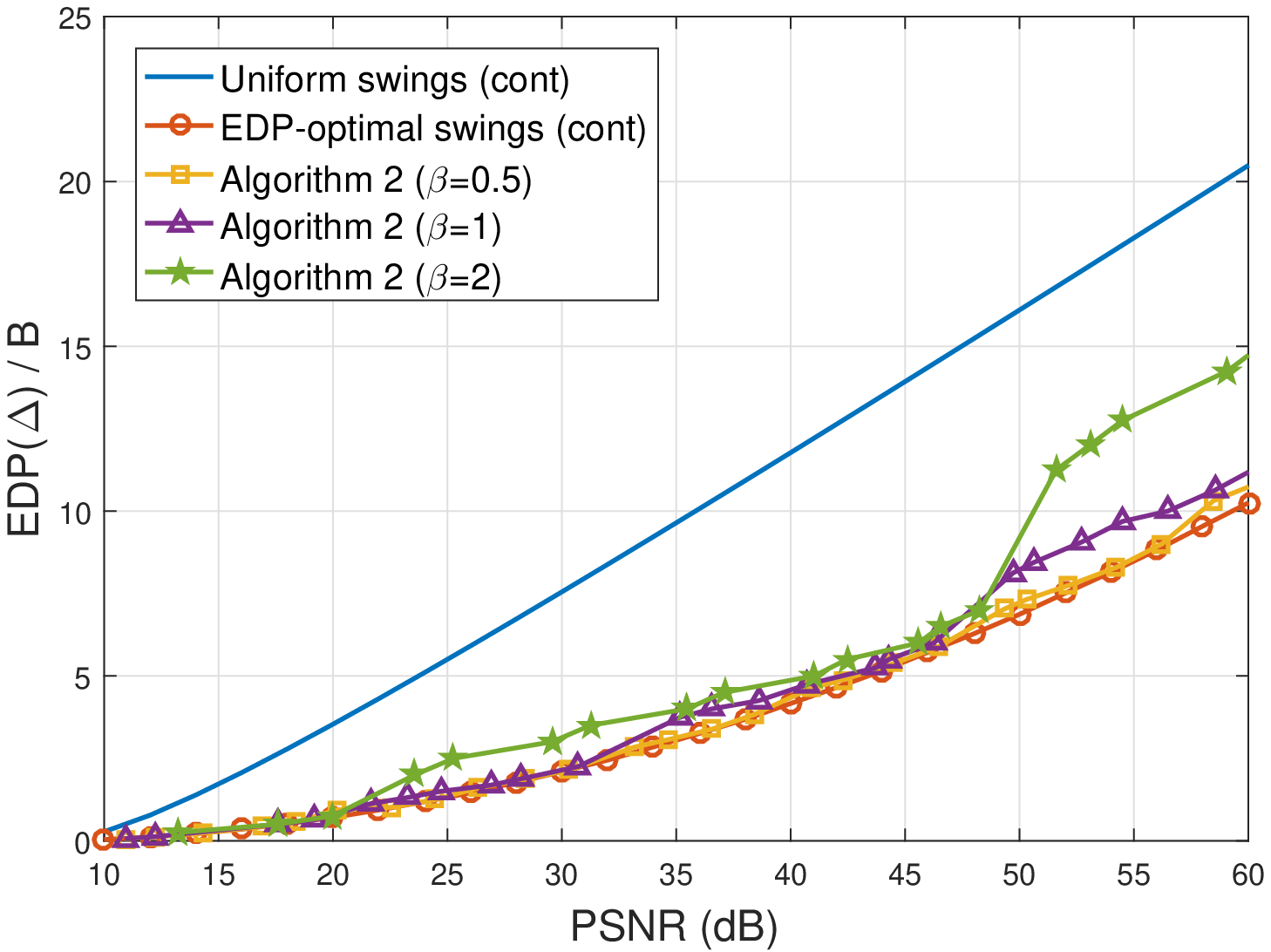}
		\label{fig:plot_edp_discrete_b}}
	\vspace{-3mm}
	\caption{EDP of discrete swings obtained by Algorithm~\ref{algo:discrete_sand_water} for (a) $B=8$ and (b) $B=16$ ($\sigma$ = 1).}
	\label{fig:plot_edp_discrete}
	\vspace{-8mm}
\end{figure}

Fig.~\ref{fig:plot_edp_discrete} compares the EDP by optimal swings of Theorem~\ref{thm:min_edp} and discrete swings by Algorithm~\ref{algo:discrete_sand_water}. By comparing Fig.~\ref{fig:plot_energy_discrete} to Fig.~\ref{fig:plot_edp_discrete}, we observe that the EDP is more sensitive to $\beta$ than the energy. The reason is that the EDP is perturbed by the discretization of $\rho$ as well as the discretization of energy. Nonetheless, the EDP penalty at PSNR = 30dB is very little for moderate granularity such as $\beta=1$. We can observe that the EDP penalty due to discrete swings is smaller for larger $B$. Since the Levin--Campello algorithm cannot solve the EDP optimization problem, it is absent in Fig.~\ref{fig:plot_edp_discrete}. 

\section{Conclusion}\label{sec:conclusion}

SRAM is a critical component for information processing systems. Casting read access for SRAMs as an end-to-end communication problem, we found the optimal bit-level swings of SRAMs for applications with fidelity dependent on bit position. We formulated convex optimization problems to determine the optimal swings for the objective functions of energy, maximum delay, and EDP. The optimized bit-level swings can achieve significant energy (50\% for 8-bit word and 75\% for 16-bit word) and EDP (45\% for 8-bit word and 75\% for 16-bit word) savings at PSNR of 30dB compared to the conventional uniform swings.  

By treating each bit position as an individual channel, we cast bit-level swing optimization problems as generalizations of water-filling that may involve sand-pouring and ground-flattening. Also, we developed optimization algorithms for discrete swings by leveraging water-filling interpretations and KKT conditions. The discrete swings obtained by proposed algorithms achieve almost the same energy and EDP savings as the continuous swings for moderate granularity. 

%As part of future work, we can consider more complicated but accurate channel models. For example, random dopant fluctuation of SRAM can be modeled by signal-dependent noise. Also, the optimization based on discretized swings due to limitations of pulse-width control units is an interesting future task. 

%% Appendix:
%% If needed a single appendix is created by
%\appendix
%% If several appendices are needed, then the command
%% in combination with further \section-commands can be used.
%\appendices

\appendices
%\section{Proof of Lemma~\ref{lemma:mse_qfunc}}\label{pf:mse_qfucn}
%The MSE of $x$ is given by
%\begin{align}
%\mathsf{MSE}(x) 
%&= \sum_{b=0}^{B-1}{4^b p_b} \label{eq:pf_mse_1} \\
%&= \sum_{b=0}^{B-1}{4^b Q\left( \frac{\Delta_b}{\sigma}\right)} \label{eq:pf_mse_2}
%\end{align}
%where \eqref{eq:pf_mse_1} was shown in \cite{Li2010maximum} and \eqref{eq:pf_mse_2} follows from \eqref{eq:ber}. 

%The decimal error $e$ is given by
%\begin{equation}\label{eq:error_representation}
%e = \sum_{b=0}^{B-1}{2^b (-1)^{q_b} e_b}
%\end{equation}
%where $q_b$ depends on $x_b$ of \eqref{eq:biterror}. If $x_b$ is random, then $q_b$ is an independent Bernoulli random variable with parameter $\frac{1}{2}$. Then, 
%\begin{align}
%&\mathsf{MSE}(x)  \\
%&= \mathbb{E}\left[ \left(\sum_{b=0}^{B-1}{2^b (-1)^{q_b} e_b}\right)^2 \right] \\
%&= \mathbb{E}\left[ \sum_{b=0}^{B-1}{4^b e_b^2} + \sum_{b_i \ne b_j}{2^{b_i + b_j} (-1)^{q_{b_i} + q_{b_j}} e_{b_i} e_{b_j}} \right] \\
%&= \sum_{b=0}^{B-1}{4^b \mathbb{E}\left[e_b^2\right]} +\sum_{b_i \ne b_j}{2^{b_i + b_j} \mathbb{E}\left[(-1)^{q_{b_i} + q_{b_j}}\right] e_{b_i} e_{b_j}} \\
%&= \sum_{b=0}^{B-1}{4^b p_b} = \sum_{b=0}^{B-1}{4^b Q\left( \frac{\Delta_b}{\sigma}\right)}
%\end{align}
%where $\mathbb{E}\left[e_b^2\right] = \mathbb{E}\left[e_b\right] = p_b$ and $\mathbb{E}\left[(-1)^{q_{b_i} + q_{b_j}}\right] = 0$. 

\section{Proof of Theorem~\ref{thm:min_energy}}\label{pf:min_energy}

The KKT conditions of \eqref{eq:min_energy} are as follows:
\begin{align}
\sum_{b=0}^{B-1}{4^b Q\left( \frac{\Delta_b}{\sigma}\right)} \le \mathcal{V}, \quad \nu &\ge 0, \label{eq:cr1_KKT_1}\\
\nu \cdot \left\{\sum_{b=0}^{B-1}{4^b Q\left( \frac{\Delta_b}{\sigma}\right)} - \mathcal{V} \right\} &= 0, \label{eq:cr1_KKT_2} \\
\Delta_b \ge 0, \quad \lambda_b \ge 0, \quad \lambda_b \Delta_b &= 0 \label{eq:cr1_KKT_3}
\end{align}
for $b \in [0, B-1]$. From $\frac{\partial L_1}{\partial \Delta_b} =0$, $\lambda_b$ is given by
\begin{equation}\label{eq:cr1_KKT_gr_1}
\lambda_b = 1 - \nu \cdot \frac{4^b}{\sqrt{2\pi} \sigma} \exp{\left(-\frac{\Delta_b^2}{2 \sigma^2}\right)} \ge 0. 
\end{equation}
By \eqref{eq:cr1_KKT_3} and \eqref{eq:cr1_KKT_gr_1}, we obtain
\begin{equation}\label{eq:cr1_KKT_slack_1}
\Delta_b \left\{1 - \nu \cdot \frac{4^b}{\sqrt{2\pi} \sigma} \exp{\left(-\frac{\Delta_b^2}{2 \sigma^2}\right)}\right\} = 0. 
\end{equation}

If $\nu = 0$, then $\lambda_b = 1$ and $\Delta_b = 0$ for any $b \in [0, B-1]$ because of \eqref{eq:cr1_KKT_3} and \eqref{eq:cr1_KKT_gr_1}. Since $\vect{\Delta} = \vect{0}$ is a trivial solution, we claim that $\nu \ne 0$, which results in
\begin{align}
\sum_{b=0}^{B-1}{4^b Q\left( \frac{\Delta_b}{\sigma}\right)} &= \mathcal{V}. 
\end{align}

If $\nu \le \frac{\sqrt{2\pi}\sigma}{4^b}$, then $\Delta_b > 0$ is impossible because it would imply $\lambda_b = 0$ and $\nu = \frac{\sqrt{2\pi}\sigma}{4^b} \exp{\left(\frac{\Delta_b^2}{2 \sigma^2}\right)}$, which contradicts the condition $\nu \le \frac{\sqrt{2\pi}\sigma}{4^b}$. Hence, $\Delta_b = 0$ for $\nu \le \frac{\sqrt{2\pi}\sigma}{4^b}$. If $\nu > \frac{\sqrt{2\pi}\sigma}{4^b}$, then $\Delta_b = 0$ is impossible because it would imply $\nu = \frac{\sqrt{2\pi}\sigma}{4^b}\exp{\left(\frac{\Delta_b^2}{2 \sigma^2}\right)} =\frac{\sqrt{2\pi}\sigma}{4^b}$, which contradicts the condition $\nu > \frac{\sqrt{2\pi}\sigma}{4^b}$. We claim that $\Delta_b > 0$ and $\lambda_b = 0$, which results in and \eqref{eq:min_energy_relation} for $\nu > \frac{\sqrt{2\pi}\sigma}{4^b}$. Thus, the optimal solution $\vect{\Delta}^*$ of \eqref{eq:min_energy} can be derived from \eqref{eq:min_energy_sol}. 

\section{Proof of Theorem~\ref{thm:max_speed}}\label{pf:max_speed}

The KKT conditions of \eqref{eq:max_speed} are as follows:
%\begin{align}
%\sum_{b=0}^{B-1}{4^b Q\left( \frac{\Delta_b}{\sigma}\right)} \le \mathcal{V}, \quad \nu &\ge 0, \label{eq:cr2_KKT_1} \\
%\nu \cdot \left\{\sum_{b=0}^{B-1}{4^b Q\left( \frac{\Delta_b}{\sigma}\right)} - \mathcal{V} \right\} &= 0, \label{eq:cr2_KKT_2} \\
%0 \le \Delta_b \le \xi, \quad \lambda_b \ge 0, \quad \lambda_b \Delta_b &= 0, \label{eq:cr2_KKT_3} \\
%\eta_b \ge 0, \quad \eta_b (\Delta_b - \xi) &= 0 \label{eq:cr2_KKT_4} 
%\end{align}
\begin{align}
\sum_{b=0}^{B-1}{4^b Q\left( \frac{\Delta_b}{\sigma}\right)} \le \mathcal{V}, \quad \nu &\ge 0, \label{eq:cr2_KKT_1} \\
\nu \cdot \left\{\sum_{b=0}^{B-1}{4^b Q\left( \frac{\Delta_b}{\sigma}\right)} - \mathcal{V} \right\} &= 0, \label{eq:cr2_KKT_2} \\
0 \le \Delta_b \le \xi, \quad \lambda_b \ge 0, \quad \lambda_b \Delta_b = 0, \quad 
\eta_b &\ge 0, \quad \eta_b (\Delta_b - \xi) = 0 \label{eq:cr2_KKT_4} 
\end{align}
for $b \in [0, B-1]$. From $\frac{\partial L_2}{\partial \Delta_b} = 0$ and $\frac{\partial L_2}{\partial \xi} = 0$, we obtain the following equations:
\begin{align}
\lambda_b = \eta_b - \nu \cdot \frac{4^b}{\sqrt{2\pi} \sigma} \exp{\left(-\frac{\Delta_b^2}{2 \sigma^2}\right)} &\ge 0,\label{eq:cr2_KKT_gr_1} \\ 
\sum_{b=0}^{B-1}{\eta_b} &= 1 \label{eq:cr2_KKT_gr_2}
\end{align}

From \eqref{eq:cr2_KKT_4} and \eqref{eq:cr2_KKT_gr_1}, 
\begin{equation} \label{eq:cr2_KKT_temp1}
\left\{\eta_b - \nu \cdot \frac{4^b}{\sqrt{2\pi} \sigma} \exp{\left(-\frac{\Delta_b^2}{2 \sigma^2}\right)} \right\} \Delta_b = 0. 
\end{equation}
If $\nu = 0$, thIEen $\eta_b \Delta_b = 0$. Also, note that $\eta_b (\Delta_b - \xi) = 0$ from \eqref{eq:cr2_KKT_4}. Both $\eta_b \Delta_b = 0$ and $\eta_b (\Delta_b - \xi) = 0$ result in $\eta_b = 0$ for any $b \in [0, B-1]$, which violates \eqref{eq:cr2_KKT_gr_2}. Hence, we claim that  
\begin{equation} \label{eq:cr2_nu}
\nu > 0, \quad \sum_{b=0}^{B-1}{4^b Q\left( \frac{\Delta_b}{\sigma}\right)} = \mathcal{V}.
\end{equation}

From~\eqref{eq:cr2_KKT_gr_1}, $\nu \le \eta_b \cdot \frac{\sqrt{2\pi} \sigma}{4^b} \exp{\left(\frac{\Delta_b^2}{2 \sigma^2}\right)}$. If $\nu\le \eta_b \cdot \frac{\sqrt{2\pi} \sigma}{4^b}$, then $\Delta_b = 0$ and $\eta_b = 0$, which violates $\nu > 0$ of \eqref{eq:cr2_nu}. Hence, $\nu > \eta_b \cdot \frac{\sqrt{2\pi} \sigma}{4^b}$, which implies $\Delta_b > 0$ and $\lambda_b = 0$ for all $b \in [0, B-1]$ because of \eqref{eq:cr2_KKT_4}. By $\lambda_b = 0$ and \eqref{eq:cr2_KKT_gr_1}, 
\begin{equation} \label{eq:cr2_eta}
\eta_b = \nu \cdot \frac{4^b}{\sqrt{2\pi} \sigma} \exp{\left(-\frac{\Delta_b^2}{2 \sigma^2}\right)}. 
\end{equation}
Because of $\nu > 0$ and \eqref{eq:cr2_KKT_4}, we claim that $\eta_b>0$ and
\begin{equation} \label{eq:cr2_uniform}
\Delta_b = \xi 
\end{equation}
for all $b \in [0, B-1]$. Hence, the optimal solution of \eqref{eq:max_speed} is \emph{uniform} swings, i.e., $\vect{\Delta}^* = (\xi, \ldots, \xi)$ where $\rho = \max\left(\vect{\Delta}^*\right) = \xi$. We confirm that the reformulated problem \eqref{eq:max_speed} is equivalent to the original problem \eqref{eq:max_speed0}. 

By \eqref{eq:cr2_eta} and \eqref{eq:cr2_uniform}, 
\begin{equation} \label{eq:cr2_relation}
\nu = \frac{\sqrt{2\pi} \sigma}{4^b} \cdot \eta_b \cdot \exp{\left(\frac{\rho^2}{2 \sigma^2}\right)} 
\end{equation}
which is equivalent to \eqref{eq:max_speed_relation}. From \eqref{eq:cr2_KKT_gr_2} and \eqref{eq:cr2_relation}, we obtain \eqref{eq:max_speed_sol} and \eqref{eq:max_speed_eta}. 

\section{Proof of Theorem~\ref{thm:min_edp} and Corollary~\ref{cor:sand_depth}}\label{pf:min_edp}

The KKT conditions of \eqref{eq:min_edp} are as follows:
%\begin{align}
%\sum_{b=0}^{B-1}{4^b Q\left( \frac{\Delta_b}{\sigma}\right)} \le \mathcal{V}, \quad \nu &\ge 0, \label{eq:cr3_KKT_1} \\
%\nu \cdot \left\{\sum_{b=0}^{B-1}{4^b Q\left( \frac{\Delta_b}{\sigma}\right)} - \mathcal{V} \right\} &= 0, \label{eq:cr3_KKT_2} \\
%0 \le \Delta_b \le \xi, \quad \lambda_b \ge 0, \quad \lambda_b \Delta_b &= 0, \label{eq:cr3_KKT_3} \\
%\eta_b \ge 0, \quad \eta_b (\Delta_b - \xi) &= 0 \label{eq:cr3_KKT_4} 
%\end{align}
\begin{align}
\sum_{b=0}^{B-1}{4^b Q\left( \frac{\Delta_b}{\sigma}\right)} \le \mathcal{V}, \quad \nu &\ge 0, \label{eq:cr3_KKT_1} \\
\nu \cdot \left\{\sum_{b=0}^{B-1}{4^b Q\left( \frac{\Delta_b}{\sigma}\right)} - \mathcal{V} \right\} &= 0, \label{eq:cr3_KKT_2} \\
0 \le \Delta_b \le \xi, \quad \lambda_b \ge 0, \quad \lambda_b \Delta_b = 0, \quad
\eta_b &\ge 0, \quad \eta_b (\Delta_b - \xi) = 0 \label{eq:cr3_KKT_4} 
\end{align}
for all $b \in [0, B-1]$. From $\frac{\partial L_3}{\partial \Delta_b} = 0$ and $\frac{\partial L_3}{\partial \xi} = 0$, we obtain the following equations:
\begin{align}
\xi + \eta_b   &= \lambda_b + \nu \cdot \frac{4^b}{\sqrt{2\pi} \sigma} \exp{\left(-\frac{\Delta_b^2}{2 \sigma^2}\right)},\label{eq:cr3_KKT_gr_1} \\ 
\sum_{b=0}^{B-1}{\Delta_b} &= \sum_{b=0}^{B-1}{\eta_b} \label{eq:cr3_KKT_gr_2}
\end{align}

Suppose that $\nu=0$, then $\xi + \eta_b = \lambda_b$ for all $b \in [0, B-1]$, which implies $\left( \xi + \eta_b \right) \Delta_b = 0$ because of \eqref{eq:cr3_KKT_4}. For $b$ such that $\Delta_b \ne 0$, $\eta_b = 0$ because of $\xi + \eta_b = 0$, $\eta_b \ge 0$ and $\xi \ge 0$. For $b$ such that $\Delta_b = 0$, $\eta_b = 0$ because of \eqref{eq:cr3_KKT_4}. Hence, if $\nu = 0$, then $\eta_b = 0$ for all $b\in[0, B-1]$, which implies $\Delta_b=0$ for all $b \in [0, B-1]$ due to $\Delta_b \ge 0$ and \eqref{eq:cr3_KKT_gr_2}. Thus, we claim that
\begin{equation} \label{eq:cr3_nu}
\nu > 0, \quad \sum_{b=0}^{B-1}{4^b Q\left( \frac{\Delta_b}{\sigma}\right)} = \mathcal{V}
\end{equation}
which is the same as \eqref{eq:cr2_nu}. 

By \eqref{eq:cr3_KKT_4} and \eqref{eq:cr3_KKT_gr_1},
\begin{equation}
\lambda_b \Delta_b = \nu \left\{\frac{\xi + \eta_b}{\nu} - \frac{4^b}{\sqrt{2\pi} \sigma} \exp{\left(-\frac{\Delta_b^2}{2 \sigma^2}\right)} \right\} \Delta_b = 0 
\end{equation}  
where $\frac{\nu}{\xi + \eta_b} \le \frac{\sqrt{2\pi}\sigma}{4^b} \exp{\left(\frac{\Delta_b^2}{2 \sigma^2}\right)}$ because of $\lambda_b \ge 0$. If $\frac{\nu}{\xi + \eta_b} \le \frac{\sqrt{2\pi}\sigma}{4^b}$, then $\Delta_b = 0$, which implies $\eta_b = 0$ by \eqref{eq:cr3_KKT_4}. Hence, we claim that
\begin{equation}\label{eq:cr3_sol1}
\Delta_b = 0,\quad \eta_b = 0, \quad \text{if}~ \frac{\nu}{\xi} \le \frac{\sqrt{2\pi}\sigma}{4^b}.
\end{equation}

If $\frac{\nu}{\xi + \eta_b} > \frac{\sqrt{2\pi}\sigma}{4^b}$, then $\Delta_b > 0$ and 
%If $\Delta_b > 0$, then
\begin{equation}\label{eq:cr3_sol3a}
\frac{\nu}{\xi + \eta_b} = \frac{\sqrt{2\pi}\sigma}{4^b}\exp{\left(\frac{\Delta_b^2}{2 \sigma^2}\right)}.
\end{equation}
By \eqref{eq:cr3_KKT_4} and \eqref{eq:cr3_KKT_gr_1},
\begin{align}
\eta_b (\Delta_b - \xi) = \nu \left\{\frac{4^b}{\sqrt{2\pi} \sigma} \exp{\left(-\frac{\Delta_b^2}{2 \sigma^2}\right)}- \frac{\xi - \lambda_b}{\nu} \right\} (\Delta_b - \xi) = 0 
\end{align}  
where $\frac{\nu}{\xi - \lambda_b} \ge \frac{\sqrt{2\pi}\sigma}{4^b}\exp{\left(\frac{\Delta_b^2}{2 \sigma^2}\right)}$ because of $\eta_b \ge 0$. If $\frac{\nu}{\xi - \lambda_b} \ge \frac{\sqrt{2\pi}\sigma}{4^b}\exp{\left(\frac{\xi^2}{2 \sigma^2}\right)}$, then $\Delta_b = \xi >0$, which implies $\lambda_b = 0$ by \eqref{eq:cr3_KKT_4}. Hence, we claim that 
\begin{equation}\label{eq:cr3_sol2}
\Delta_b = \xi, \quad \lambda_b = 0, \quad \text{if}~\frac{\nu}{\xi}\ge\frac{\sqrt{2\pi}\sigma}{4^b}\exp{\left(\frac{\xi^2}{2 \sigma^2}\right)}. 
\end{equation}
If $\frac{\sqrt{2\pi}\sigma}{4^b} \le \frac{\nu}{\xi - \lambda_b} < \frac{\sqrt{2\pi}\sigma}{4^b}\exp{\left(\frac{\xi^2}{2 \sigma^2}\right)}$, then 
%If $\Delta_b < \xi$, then 
\begin{equation}\label{eq:cr3_sol3b}
\frac{\nu}{\xi - \lambda_b} = \frac{\sqrt{2\pi}\sigma}{4^b}\exp{\left(\frac{\Delta_b^2}{2 \sigma^2}\right)}. 
\end{equation}
By \eqref{eq:cr3_sol3a} and \eqref{eq:cr3_sol3b}, 
\begin{equation}
\frac{\nu}{\xi + \eta_b} = \frac{\nu}{\xi - \lambda_b} = \frac{\sqrt{2\pi}\sigma}{4^b}\exp{\left(\frac{\Delta_b^2}{2 \sigma^2}\right)} 
\end{equation}
for $0 < \Delta_b < \xi$. $\xi + \eta_b = \xi - \lambda_b$ (i.e., $\eta_b = - \lambda_b$) means $\eta_b = \lambda_b = 0$ because of $\eta_b \ge 0$ and $\lambda_b \ge 0$. Hence, we claim that
\begin{align}\label{eq:cr3_sol3}
\frac{\nu}{\xi} = \frac{\sqrt{2\pi}\sigma}{4^b}\exp{\left(\frac{\Delta_b^2}{2 \sigma^2}\right)} , \quad \eta_b = \lambda_b = 0
\end{align}
for $\frac{\sqrt{2\pi}\sigma}{4^b} < \frac{\nu}{\xi} < \frac{\sqrt{2\pi}\sigma}{4^b}\exp{\left(\frac{\xi^2}{2 \sigma^2}\right)}$. 

Due to \eqref{eq:cr3_KKT_gr_2}, there should exist $\eta_b > 0$ for $b \in [0, B-1]$ to make $\sum_{b=0}^{B-1}{\Delta_b} > 0$. Hence, there exists $\Delta_b=\xi$ due to \eqref{eq:cr3_KKT_4}, which implies $\rho = \max(\vect{\Delta}) = \xi$. From \eqref{eq:cr3_sol1}, \eqref{eq:cr3_sol2}, \eqref{eq:cr3_sol3}, and $\rho = \xi$, we can obtain the optimal solution $\vect{\Delta}^{*}$ of \eqref{eq:min_edp_sol}. 

Note that $s_b > 0$ for $\Delta_b = \rho$ and $\lambda_b = 0$. In this case, \eqref{eq:cr3_KKT_gr_1} can be modified into
\begin{equation} \label{eq:cr3_sand_depth_temp0}
\rho + \eta_b = \nu \cdot \frac{4^b}{\sqrt{2\pi} \sigma} \exp{\left(-\frac{\rho^2}{2 \sigma^2}\right)}. 
\end{equation}
As shown in Fig.~\ref{fig:min_edp}\subref{fig:min_edp_a}, the sand depth $s_b$ is given by
\begin{align}
s_b &= \log{\frac{\nu}{\rho}} - \left(\log{\frac{\sqrt{2\pi}\sigma}{4^b}} + \frac{\rho^2}{2 \sigma^2}\right)= \log{\frac{\nu}{\rho}} - \log{\frac{\nu}{\rho + \eta_b}} = \log{\left(1 + \frac{\eta_b}{\rho}\right)} \label{eq:cr_3_sand_temp_temp2}
\end{align}
where \eqref{eq:cr_3_sand_temp_temp2} follows from \eqref{eq:cr3_sand_depth_temp0}. If $0 \le \Delta_b < \rho$, then $\eta_b =0$ as shown in \eqref{eq:cr3_sol1} and \eqref{eq:cr3_sol3}. Hence, $s_b = 0$ for $0 \le \Delta_b < \rho$. Hence, \eqref{eq:sand_depth} in Corollary~\ref{cor:sand_depth} is proved. Also, \eqref{eq:sand_depth_amount} in Corollary~\ref{cor:sand_depth} is derived from \eqref{eq:cr3_KKT_gr_2} and \eqref{eq:cr_3_sand_temp_temp2}. 

%\appendix

%%% Use \section* for acknowledgement
%\section*{Acknowledgment}
%
%The authors would like to thank various sponsors for supporting
%their research.

%% References:
%% We recommend the usage of BibTeX:
%%

%\IEEEtriggeratref{17}

\bibliographystyle{IEEEtran}
\bibliography{IEEEabrv,mybib}

\end{document}